\PassOptionsToPackage{table}{xcolor}
\documentclass[letterpaper,twocolumn,10pt]{article}

\usepackage{hegroup}
\usepackage[table]{xcolor}
\usepackage[normalem]{ulem}
\useunder{\uline}{\ul}{}
\usepackage{tikz}
\usepackage{amsfonts}
\usepackage{xspace} 
\usepackage{amsmath}
\usepackage{mathrsfs}
\usepackage{amssymb}
\usepackage{enumitem}
\usepackage{amsmath}
\usepackage{amsthm}
\usepackage{multirow}
\usepackage{makecell}
\usepackage{caption}
\usepackage{booktabs}
\usepackage{balance}
\usepackage{tcolorbox}
\usepackage{xurl}
\usepackage{algorithm}
\usepackage{algorithmic}
\usepackage{hyperref}
\usepackage{tabularx}
\usepackage{cleveref}

\newcommand{\mypara}[1]{\noindent{\bf {#1}.}}
\newcommand{\method}{\textbf{\textit{TH-Bench}}\xspace}
\newcommand{\attack}{\textbf{\textit{Quality-Preserving Attack}}\xspace}
\begin{document}
\pagestyle{plain}
\title{\method: 
Evaluating Evading Attacks via Humanizing AI Text on Machine-Generated Text Detectors}
\date{}
\author{
Jingyi Zheng\thanks{Equal contribution.}     \ \ \
Junfeng Wang\footnotemark[1]   \ \ \ 
Zhen Sun \ \ \ 
Wenhan Dong  \ \ \ 
Yule Liu  \ \ \ 
Xinlei He\thanks{Corresponding author (\href{mailto:xinleihe@hkust-gz.edu.cn}{xinleihe@hkust-gz.edu.cn}). } \ \ \
\\
\\
\textit{The Hong Kong University of Science and Technology (Guangzhou)} 
}

\maketitle
\begin{abstract}

As Large Language Models (LLMs) advance, Machine-Generated Texts (MGTs) have become increasingly fluent, high-quality, and informative. 
Existing wide-range MGT detectors are designed to identify MGTs to prevent the spread of plagiarism and misinformation.
However, adversaries attempt to humanize MGTs to evade detection (named evading attacks), which requires only minor modifications to bypass MGT detectors.
Unfortunately, existing attacks generally lack a unified and comprehensive evaluation framework, as they are assessed using different experimental settings, model architectures, and datasets.
To fill this gap, we introduce the Text-Humanization Benchmark (\method), the first comprehensive benchmark to evaluate evading attacks against MGT detectors.
\method evaluates attacks across three key dimensions: evading effectiveness, text quality, and computational overhead.
Our extensive experiments evaluate 6 state-of-the-art attacks against 13 MGT detectors across 6 datasets, spanning 19 domains and generated by 11 widely used LLMs. 
Our findings reveal that no single evading attack excels across all three dimensions.
Through in-depth analysis, we highlight the strengths and limitations of different attacks.
More importantly, we identify a trade-off among three dimensions and propose two optimization insights. 
Through preliminary experiments, we validate their correctness and effectiveness, offering potential directions for future research.

\end{abstract}

\begin{figure*}
    \centering
    \includegraphics[width=\textwidth]{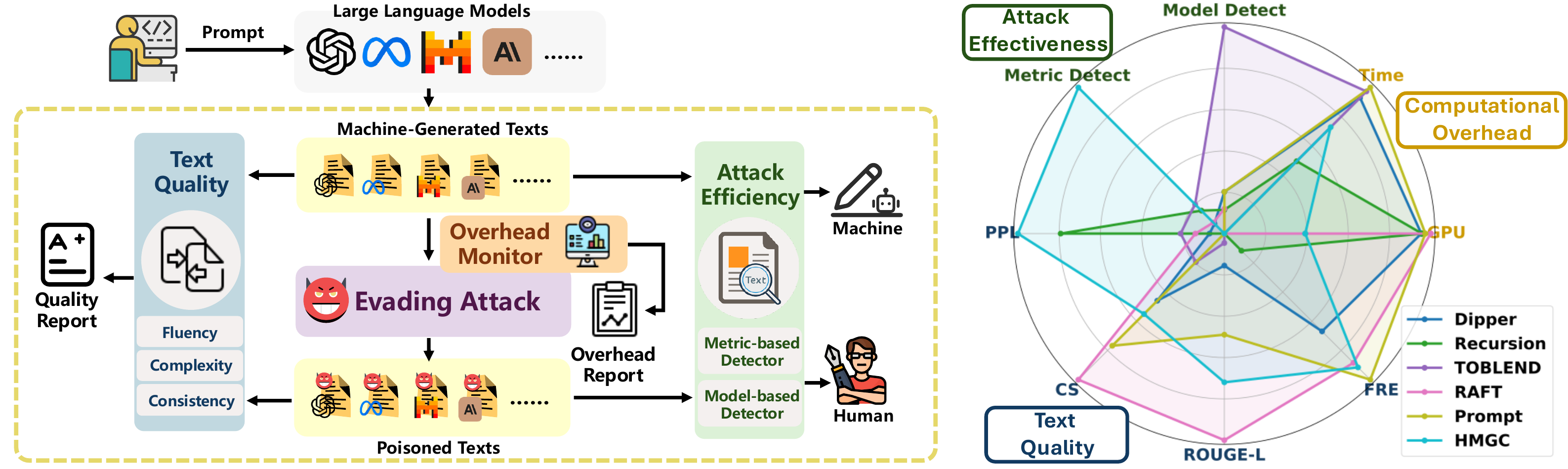} 
    \caption{\method benchmarks evading attacks in terms of effectiveness against MGT detectors, impact on text quality, and computational overhead. The right figure shows the normalized results of various attacks we evaluate across three dimensions. A higher value for each point indicates better performance of the attack in the corresponding metric.}
    \label{fig:fig1} 
\end{figure*}

\section{Introduction}

Recently, remarkable Large Language Models (LLMs) such as GPT~\cite{achiam2023gpt}, PaLM~\cite{anil2023palm}, and Llama~\cite{touvron2023llama} have exhibited exceptional abilities across diverse fields, such as physics~\cite{west2023ai}, medicine~\cite{digiorgio2023artificial}, mathematics~\cite{li2023solving}, and linguistics~\cite{liu2023summary}.
In natural language processing (NLP), LLMs showcase to produce high-quality, fluent, and knowledge-rich text, which could tackle a wide range of tasks. 
Despite these advancements, hallucination~\cite{mckenna2023sources}, misinformation dissemination~\cite{bian2024influence}, and societal bias~\cite{ferrara2023should} are still serious challenges that LLMs are facing.  
Thus, detecting the misuse of Machine-Generated Texts (MGTs) has become a critical research priority~\cite{sun2024we}. 

Recent studies introduced several kinds of detectors to identify MGTs.
These detectors can be divided into two categories, metric-based and model-based. 
The metric-based detectors~\cite{gehrmann2019gltr,mitchell2023detectgpt,su2023detectllm} extract text features by using proxy LLMs and then train a binary classifier to distinguish the MGTs. 
The model-based detectors~\cite{ippolito2019automatic,guo2023close} modify LLMs like BERT~\cite{devlin2018bert} by adding a classification head and then fine-tuning the whole model on the MGT dataset. 
Several benchmarks have demonstrated the effectiveness of these MGT detectors, like MGTBench~\cite{he2024mgtbench} and DetectRL~\cite{wu2024detectrl}. 
However, recent studies~\cite{wu2025survey, yang2023survey, wu2024detectrl} show that humanizing MGTs could serve as a form of evading attacks to bypass detection.
Current evading attacks can be divided into three categories including paraphrase, perturbation, and data mixing attacks.
Paraphrase attacks~\cite{krishna2024paraphrasing} focus on rewriting text while maintaining its original meaning to evade detection. 
Perturbation attacks~\cite{wang2024raft} introduce subtle modifications to the generated text, simulating common writing errors or word substitutions.
Data mixing attacks~\cite{huang2024toblendtokenlevelblendingensemble} combine texts from multiple sources, such as mixing the texts generated by multiple LLMs, to disrupt the detector’s ability. 
Among them, some attacks~\cite{ren2019generating,jin2020bert} require only simple modifications to bypass MGT detectors.
These low-cost attacks could have serious consequences, such as allowing MGTs to evade regulation and undermining trust in MGT detectors.
Unfortunately, existing attack methods against MGT detectors generally assess using different experimental settings, model architectures, and datasets, thus needing a comprehensive benchmark to quantify the attacks from various dimensions.

To fill this gap, in this paper, we study the following  questions:
\textbf{ (1) How effective are these evading attacks against different MGT detectors? 
(2) To what extent do these evading attacks impact the quality of the MGTs? 
(3) What is the overhead to run these attacks? } 
To answer these questions, we introduce the Text-Humanization Benchmark (\method), a novel benchmark designed to evaluate evading attacks against MGT detectors. 
The overall framework is shown in~\Cref{fig:fig1}.
For the first question, we consider 13 representative MGT detectors, encompassing both metric-based and model-based to evaluate the effectiveness of 6 state-of-the-art evading attacks which cover three categories (paraphrasing, perturbations, and data mixing attacks). 
The evaluation is carried out on 6 datasets~\cite{he2024mgtbench,liu2024generalization}, containing 19 domains of MGTs generated by 11 widely used LLMs.
We conduct comprehensive evaluations of the effectiveness of these attacks on both binary classification tasks, determining whether a text is MGT or not, and multiclass classification tasks, focusing on text attribution.
Furthermore, we also evaluate the effectiveness of these evading attacks against detectors that have been adapted to known evading attacks.
For the second question, we evaluate text quality from three perspectives: text fluency; text semantic consistency, which quantifies the similarity between texts before and after the attack; and text complexity.
For the third question, we track the computational overhead of different attacks by measuring their execution time and GPU memory consumption across varying token lengths. 
After extensive experiments, we conduct an in-depth analysis of the strengths and weaknesses of the evaluated attacks, as well as the challenges they face in real-world applications.
The results of experiments reveal that no single evading attack excels across all three dimensions. 
More importantly, based on our findings, we identify a trade-off among evading effectiveness, text quality, and computational cost.
Furthermore, we propose two optimization insights called Quality Preserving Attack and Attack Blending to improve the trade-off, which have validated their effectiveness through preliminary experiments.

In summary, we make the following contributions:

\begin{itemize}[leftmargin=*]
    \item We propose \method, the first comprehensive benchmark that evaluates text humanization evading attacks against various MGT detection methods in three dimensions.
    \item We conduct extensive experiments to evaluate various state-of-the-art evading attacks and provide an in-depth analysis of their strengths, weaknesses, and challenges in real-world applications. 
    \item 
    We identify a trade-off and two optimization insights, validating their effectiveness and correctness through experiments, which provide directions for future work.
\end{itemize}

\section{Preliminary}

\subsection{MGT Detection}
\label{subsec:mgt-detection}

Current detection methods can be broadly classified into two categories based on their design: metric-based and model-based methods~\cite{liu2024generalization,he2024mgtbench}.
Metric-based methods predict the origin of a text by analyzing measurable features such as textual consistency, complexity, word rank, and entropy. 
This method typically uses pre-trained language models to extract contextual features, which are later classified by a binary classifier ~\cite{gehrmann2019gltr,mitchell2023detectgpt,su2023detectllm}. 
In this work, we consider the following metric-based methods. 
\textbf{Log-Likelihood~\cite{solaiman2019release}} averages the token-wise log probabilities, with a higher score indicating a greater likelihood of being MGT. 
\textbf{Rank~\cite{gehrmann2019gltr}} calculates the average absolute rank of each word, with a smaller score indicating a greater MGT likelihood of being MGT. 
\textbf{Log-Rank~\cite{mitchell2023detectgpt}} applies the logarithm to the rank of each word and averages these values, with a lower score indicating a greater MGT likelihood. 
\textbf{Entropy~\cite{gehrmann2019gltr}} calculates the average entropy of each word conditioned on its previous context, with a lower score indicating a greater MGT likelihood. 
\textbf{GLTR~\cite{gehrmann2019gltr}} is a tool designed to assist in identifying MGT, and in our evaluation, following~\cite{guo2023close}, we use the Test-2 features, which focus on the fraction of words ranked within 10, 100, 1,000, and others. 
\textbf{LRR~\cite{su2023detectllm}} combines Log-Likelihood and Log-Rank, with a higher score indicating a greater MGT likelihood. 
\textbf{Fast-DetectGPT~\cite{bao2023fast}} is a detector based on DetectGPT~\cite{gehrmann2019gltr} by replacing perturbation with efficient sampling to improve. 
Following the authors’ optimal settings, we use \texttt{GPT-Neo-2.7b} as the scoring model and \texttt{GPT-J-6b} as the reference model. 
\textbf{Binoculars~\cite{hans2024spotting}} calculates the perplexity-to-cross-perplexity ratio using two LLMs, with a lower score indicating a greater MGT likelihood.
Following the authors’ optimal settings, we used \texttt{Falcon-7B-Instruct} for PPL and \texttt{Falcon-7B} with \texttt{Falcon-7B-Instruct} for X-PPL.

In contrast, model-based methods often rely on training models. 
One common approach is to add a classification layer to the BERT model and fine-tune it on a dataset containing labeled examples of both human-written and machine-generated text, resulting in a classification model that can perform text source judgment~\cite{tao2024cudrt,ippolito2019automatic,guo2023close}. 
In this work, we consider the following model-based methods. 
\textbf{RADAR~\cite{hu2023radar}} employs adversarial training between a paraphraser and a detector, with the pre-trained RoBERTa detector from Hugging Face used without further training in our work. 
\textbf{OpenAI Detector~\cite{solaiman2019release}} is a RoBERTa-based model fine-tuned on outputs from the largest GPT-2 model (1.5B parameters) to predict whether a given text is MGT or not. 
\textbf{ChatGPT Detector~\cite{guo2023close}} is created by fine-tuning a RoBERTa model using the HC3~\cite{guo2023close} dataset, using only the answered text in our evaluation for consistency with other methods.
\textbf{Contrastive Domain Adaptation (ConDA)~\cite{bhattacharjee2023conda}} uses contrastive learning to acquire domain-invariant representations, enhancing the model’s ability to generalize across different domains.
\textbf{LM Detector~\cite{he2024mgtbench}} is built by fine-tuning the pre-trained language model (LM) with an extra classification layer. 

\subsection{Evading Attack}
\label{subsec:adv}

A few recent works have attempted to attack text detection models. 
Following the recent studies~\cite{wu2025survey,yang2023survey,wu2024detectrl}, the evading attacks can be broadly classified into three categories: paraphrase, perturbation, and data mixing attacks. 

Paraphrase attacks have been widely studied as an effective strategy to evade MGTs detection~\cite{wu2024detectrl}. 
The core idea is rewriting text while maintaining its original meaning. 
In this work, we consider the following specific types of paraphrase attacks: 
Krishna et al.~\cite{krishna2024paraphrasing} build an 11B parameter paraphrase generation model that can paraphrase paragraphs, condition on the surrounding context, and control lexical diversity and content reordering which is called \textbf{Dipper}. 
Sadasivan et al.~\cite{sadasivan2023can} propose \textbf{Recursion}, which builds on paraphrasers such as Dipper, \texttt{LLaMA-2-7B-Chat}~\cite{touvron2023llama} and 222M T5-based paraphraser~\cite{damodaran2021parrot} .
They incorporate recursion into evading attack by iteratively feeding the original MGT into the paraphraser. 
The output of the paraphraser is then reintroduced as input for subsequent iterations. 
This process is repeated multiple times, progressively altering the text and making it increasingly difficult for the detector to identify MGT. 
Leveraging crafted prompts to guide LLMs into generating text that closely resembles human-authored content is a paraphrasing method that is commonly utilized in everyday scenarios.
We adopt \textbf{Prompt~\cite{xu2023llm}}, consists of three key components: (1) Original Input, which includes the original sample along with its ground-truth label, serving as the foundation for the attack; (2) Attack Objective, which describes the task of generating a modified sample that retains the original semantic meaning while deceiving the detection model; and (3) Attack Guidance, which provides rephrase instructions to guide the LLMs in altering the original sample at different linguistic levels. 

Perturbation attacks focus on introducing adversarial perturbations on the MGT.
These perturbations intend to mimic human writing patterns, which tend to be writing errors, word substitutions, or other noises that are common in the writing process. 
In this work, we consider the following specific types of perturbation attacks: 
\textbf{RAFT~\cite{wang2024raft}} is a zero-shot black-box attack framework that leverages an auxiliary LLM embedding to select words in MGTs for substitution. 
It then uses a black-box LLM to generate candidate replacements, greedily selecting the one that most effectively bypasses the detectors.
\textbf{HMGC~\cite{zhou2024humanizing}} determines the importance of words in the original text by first training a surrogate detection model and then modifying the text by synonym replacement. Compared to RAFT, HMGC reduces resource overhead during word substitution. However, it requires the user to gather a dataset similar to the attacked text and train a surrogate model, which can be resource-intensive and costly. 

Data mixing attacks refer to a multi-LLM mixing attack, which creates MGTs by sampling and combining sentences from multiple LLMs. 
In our work, we adopt \textbf{TOBLEND~\cite{huang2024toblendtokenlevelblendingensemble}}, a token-level ensemble text generation method that constructs new sentences by randomly sampling tokens from a set of candidate LLMs.

\begin{table*}[htbp]
\setlength{\tabcolsep}{6pt} 
\renewcommand{\arraystretch}{0.9} 
\centering
\caption{AUC of evading attacks across detectors and datasets, averaged over LLMs, in binary classification task. OpenAI-D, ChatGPT-D, LM-D, HMGC, and HMGC(Mis) represent the OpenAI, ChatGPT, LM Detectors, Standard HMGC, and Mismatched HMGC respectively. Attack effectiveness is assessed by the AUC difference from the Clean AUC, with a larger difference indicating greater effectiveness.}
\label{tab:big}
\resizebox{0.98\textwidth}{!}{
\begin{tabular}{c|c|c|c|c|c|c|c|c|c|c|c|c|c|c}
    \toprule
    Dataset&\makecell{Detector$\rightarrow$\\ Method $\downarrow$}& \makecell{Log-\\Likelihood} & Rank & \makecell{Log-\\Rank} & Entropy & GLTR & \makecell{Bino-\\culars} & LRR & RADAR & \makecell{Open\\AI-D} & \makecell{Chat\\GPT-D} & \makecell{LM\\-D} & ConDA & \makecell{Fast-De-\\tect-GPT}\\ 
    \midrule
    &\makecell{Clean} &0.913&0.798&0.918&0.799&0.924&0.986&0.916&0.947&0.778&0.724&0.999&0.338&0.941   \\
    \cmidrule{2-15} 
    &\makecell{Dipper} & 0.832&0.714&0.841&0.562&0.858&0.988&0.841&\textbf{0.913}&0.895&0.772&0.991&0.304 &0.961  \\ 
    \multirow{3}{*}{\rotatebox{90}{Essay}}&\makecell{TOBLEND}  & 0.508&0.658&0.620&0.444&0.618&0.744&0.848&0.951&0.642&\textbf{0.005}&\textbf{0.954}&0.225 &0.769  \\
    &\makecell{Recursion} & 0.466 &0.560&0.488&\textbf{0.216}&0.541&0.924&0.539&0.962&0.856&0.688&0.975&\textbf{0.179} &0.773   \\ 
    &\makecell{Prompt}& 0.939&0.793&0.936&0.806&0.921&0.998&0.890&0.975&0.626&0.714&0.973&0.446&0.967   \\ 
    &\makecell{HMGC}& \textbf{0.185}&\textbf{0.301}&\textbf{0.183}&0.277&\textbf{0.207}& \textbf{0.419}&\textbf{0.185}&0.933&0.853&0.518&0.998&0.602  &\textbf{0.355} \\
    &\makecell{HMGC (Mis)}& 0.900&0.774&0.905&0.776&0.911&0.983&0.892&0.954&0.769&0.719&1.000&0.342  &0.933 \\
    &\makecell{RAFT}& 0.776&0.656&0.834&0.804&0.859&0.859&0.926&0.928&\textbf{0.578}&0.694&0.998&0.382 &0.712  \\
    \midrule
    
    &\makecell{Clean} & 0.907&0.868&0.907&0.801&0.907&0.984&0.882&0.946&0.830&0.727&0.998&0.264 &0.919  \\
    \cmidrule{2-15}
    &\makecell{Dipper} & 0.776&0.751&0.791&0.596&0.795&0.967&0.807&\textbf{0.898}&0.948&0.788&0.996&0.210 &0.932   \\ 
    \multirow{3}{*}{\rotatebox{90}{WP}}&\makecell{TOBLEND}  & 0.568&0.754&0.671&0.491&0.628&0.822&0.845&0.968&0.856&\textbf{0.000}&\textbf{0.895}&0.122 &0.802  \\ 
    &\makecell{Recursion} & 0.358&0.578&0.403&0.260&0.437&0.848&0.528&0.990&0.977&0.730&0.984&\textbf{0.060} &0.659  \\ 
    &\makecell{Prompt}& 0.928&0.865&0.920&0.818&0.901&0.994&0.860&0.981&\textbf{0.759}&0.665&0.954&0.382 &0.947  \\
    &\makecell{HMGC}& \textbf{0.141}&\textbf{0.186}&\textbf{0.137}&\textbf{0.151}&\textbf{0.150}&\textbf{0.370}&\textbf{0.151}&0.968&0.913&0.436&0.993&0.594 &\textbf{0.293}  \\
    &\makecell{HMGC (Mis)}& 0.890&0.834&0.892&0.776&0.892&0.978&0.869&0.955&0.824&0.719&0.999&0.268 &0.909  \\
    &\makecell{RAFT}& 0.867&0.787&0.886&0.807&0.896&0.886&0.996&0.942&0.799&0.713&0.998&0.281 & 0.694 \\
    
    \midrule
    &\makecell{Clean} &0.808&0.714&0.821&0.674&0.863&0.980&0.871&0.998&0.932&0.812&1.000&0.199 &0.930  \\
    \cmidrule{2-15}
    &\makecell{Dipper} & 0.684&0.606&0.710&0.551&0.719&0.980&0.709&\textbf{0.997}&0.976&0.868&0.999&0.179 & 0.924  \\ 
    \multirow{3}{*}{\rotatebox{90}{Reuters}}&\makecell{TOBLEND}  &0.293&0.479&0.375&0.469&0.331&0.674&0.672&0.998&\textbf{0.813}&\textbf{0.004}&\textbf{0.990}&0.185 & 0.682  \\
    &\makecell{Recursion} & 0.300&0.438&0.327&\textbf{0.446}&0.326&0.872&0.374&1.000&0.987&0.766&0.993&\textbf{0.059} &0.655  \\
    &\makecell{Prompt}& 0.779&0.682&0.784&0.571&0.799&0.977&0.771&1.000&0.874&0.839&1.000&0.331 &0.907  \\ 
    &\makecell{HMGC}& \textbf{0.020}&\textbf{0.163}&\textbf{0.164}&0.448&\textbf{0.101}&\textbf{0.394}&\textbf{0.037}&1.000&0.986&0.460&0.997&0.659 &\textbf{0.295}  \\  
    &\makecell{HMGC (Mis)}& 0.796&0.673&0.811&0.673&0.845&0.972&0.846&0.999&0.929&0.808&1.000&0.204 &0.912  \\  &\makecell{RAFT}&0.820&0.716&0.835&0.811&0.831&0.852&0.806&0.999&0.825&0.796&1.000&0.234 &0.712  \\
    \midrule

    &\makecell{Clean} &0.671  & 0.615 &0.679  &0.586 &0.678 & 0.861 & 0.674& 0.756& 0.651&0.740 &0.956 & 0.397  & 0.828  \\
    \cmidrule{2-15}
    &\makecell{Dipper}&0.609 &0.547 &0.612  & 0.534 & 0.617&0.894 & 0.612 &0.758 &0.848 &0.670&0.820& 0.325  &0.876 \\   
    \multirow{3}{*}{\rotatebox{90}{STEM}}&\makecell{TOBLEND}  &0.478  &0.543  &0.527  &\textbf{0.511} &0.508 & 0.587 & 0.721&0.794 &0.614 &\textbf{0.004} &\textbf{0.745} & 0.339  &0.675  \\ 
    &\makecell{Recursion} & 0.562 & 0.523 &0.559  & 0.518& 0.568&0.883  &0.546 &0.878 &0.892 & 0.700&0.792 &\textbf{0.216}  &0.823     \\ 
    &\makecell{Prompt}& 0.654 & 0.606 & 0.659  & 0.579&0.650 & 0.844 & 0.627& 0.774& \textbf{0.568}& 0.746&0.963 &0.426   &0.784   \\
    &\makecell{HMGC}&\textbf{0.390}  &\textbf{0.323}  &\textbf{0.343}  &0.512 &\textbf{0.335} & \textbf{0.257} &\textbf{0.208} & 0.818 &0.872 &0.459 &0.920 &0.523 &\textbf{0.220}   \\ 
    &\makecell{HMGC (Mis)}&0.467  &0.441&0.455  &0.517 &0.463 & 0.835 &0.432 &0.760 &0.665&0.731&0.932 &0.400 &0.800   \\ 
    &\makecell{RAFT}&0.593  &0.563  & 0.614 &0.583 &0.615 & 0.844 &0.661 &\textbf{0.742} &0.583 &0.707 &0.953 &0.410 &0.786      \\
    \midrule
    
    &\makecell{Clean} &0.803&0.689&0.801&0.639&0.792&0.902 &0.758& 0.745&0.740&0.740&0.969&0.376 &0.870   \\
    \cmidrule{2-15}
    \multirow{3}{*}{\rotatebox{90}{Social Science}}&\makecell{Dipper}& 0.696&0.605&0.703&0.433&0.711&0.925 &0.695& \textbf{0.728}&0.882&0.680&0.805&0.334 &0.922\\ 
    &\makecell{TOBLEND} &0.538&0.666&0.649&0.484&0.609&0.658 &0.852& 0.749&0.653&\textbf{0.169}&\textbf{0.750}&0.313&0.803 \\
    &\makecell{Recursion}& 0.626&0.582&0.637&0.376&0.651&0.909 &0.634&0.832 & 0.905&0.705&0.806&\textbf{0.261} &0.876  \\
    &\makecell{Prompt}& 0.772&0.679&0.762&0.630&0.738& 0.877&0.701& 0.760&\textbf{0.643}&0.735&0.971&0.417 &0.832 \\
    &\makecell{HMGC}&  \textbf{0.130}&\textbf{0.228}&\textbf{0.135}&\textbf{0.216}&\textbf{0.158}&\textbf{0.337} &\textbf{0.182}& 0.798&0.897&0.470&0.899&0.538 & \textbf{0.297}\\ 
    &\makecell{HMGC (Mis)}&  0.407&0.376&0.410&0.380&0.426&0.876 &0.417& 0.749&0.747&0.729&0.933&0.380 & 0.844\\ &\makecell{RAFT}&0.669&0.643&0.695&0.631&0.688&0.857 &0.736& 0.767&0.647&0.717&0.972&0.392 & 0.820\\
    \midrule
    
    &\makecell{Clean} &0.774  &0.697  & 0.774 &0.636 &0.763 & 0.885 &0.737 & 0.809&0.757 &0.754 & 0.961&0.346   &0.836    \\
    \cmidrule{2-15}
    &\makecell{Dipper} & 0.666 &  0.611& 0.680 &0.420 & 0.682& 0.905 &0.688 &\textbf{0.752} &0.898 &0.684 & 0.724& 0.305 &0.888    \\ 
    \multirow{3}{*}{\rotatebox{90}{Humanity}}&\makecell{TOBLEND} & 0.497 & 0.681 &0.633  &0.407 &0.561 & 0.696 & 0.866&0.814 & \textbf{0.666}& 0.501&\textbf{0.659} &0.259 &0.813    \\ 
    &\makecell{Recursion} &0.608  & 0.592 & 0.628 &0.379 & 0.640& 0.883 &0.645 &0.841 &0.910 &0.700 &0.765 & \textbf{0.252}  &0.839    \\
    &\makecell{Prompt}&0.757&0.711&0.749&0.638&0.715& 0.870&0.690&0.830 & 0.687&0.747&0.969&0.400 &0.800\\ &\makecell{HMGC}&\textbf{0.099}&\textbf{0.140}&\textbf{0.101}&\textbf{0.121}&\textbf{0.132}&\textbf{0.299} &\textbf{0.151}& 0.851&0.889&\textbf{0.461}&0.899&0.483  &\textbf{0.277}   \\
    &\makecell{HMGC (Mis) }&0.473&0.416&0.475&0.406&0.493&0.876 &0.480&0.810 &0.759&0.752&0.945&0.347  &0.827    \\ &\makecell{RAFT}&0.598&0.596&0.635&0.569&0.629&0.881 &0.723& 0.819& 0.669&0.728&0.974&0.364  &0.818    \\
   
    \bottomrule
\end{tabular}}
\end{table*}

\section{TH-Bench}

\subsection{Task Definition}

\noindent\textbf{Task 1: Attack Effectiveness Assessment.}
This task aims to comprehensively evaluate the performance of text humanization evading attacks across different datasets, LLMs, and MGT detectors in binary classification tasks (MGT or not) and multiclass classification tasks (Text attribution).
Additionally, we assess the effectiveness of the attacks as emerging threats when the MGT detectors have already been adapted to defend against known evading attacks.

\noindent\textbf{Task 2: Text Quality Assessment.} 
Another goal of evading attack is to maintain the quality of the text. 
Therefore, the impact of the attack on text quality serves as a key factor in assessing its performance.
In this benchmark, we assess text quality across three aspects: \textit{Fluency}, \textit{Semantic Consistency}, and \textit{Complexity}. 
Ideally, the attack should introduce only minimal changes in these three aspects to preserve the original characteristics of the text.

\noindent\textbf{Task 3: Attack Execution Overhead Assessment.} 
The overhead affects the usefulness of the attack in real-world scenarios, making it critical to evaluate it when executing the attack.
We evaluate the execution overhead of each attack in terms of processing time and memory consumption in various token lengths.

\subsection{Framework}

\mypara{Datasets}
In this work, we consider six datasets from MGTBench~\cite{he2024mgtbench} and MGT-Academic~\cite{liu2024generalization}.
MGTBench consists of three datasets (Essay, WP, Reuters), each containing 1,000 articles, with MGT data generated by seven different LLMs. 
These datasets are split into training and test sets in an 80/20 ratio.
MGT-Academic covers STEM, Social Science, and Humanity datasets, comprising a total of 16 domains, with data sourced from Wiki, Arxiv, and Gutenberg and MGT data generated by five LLMs. 
The size of the training and test set are consistent with that of MGTBench.
More details about these datasets can be found in ~\Cref{sec:datasets}.

\mypara{Detectors}
We utilize 13 representative detectors, encompassing both metric-based and model-based approaches, as discussed in~\Cref{subsec:mgt-detection}. 
For metric-based methods, we use \texttt{Llama-2-7B} as the base model. 
The base model extracts relevant metrics from the text, which are subsequently used to train a logistic regression classifier for the prediction.
For model-based detectors, we directly use their officially released model weights.
Following ~\cite{he2024mgtbench},
we employ the \texttt{RoBERTa-base} version for the OpenAI Detector, as it offers better detection performance. 
For the ChatGPT Detector, we utilize the \texttt{RoBERTa-base} model provided with it. 
For the LM Detector, we adopt the distilled \texttt{BERT-base} for its balanced trade-off between performance and efficiency. 
For ConDA, we leverage the \texttt{RoBERTa-base} model, which is specifically optimized for detecting ChatGPT-generated text.

\begin{figure}
    \centering
    \includegraphics[width=\columnwidth]{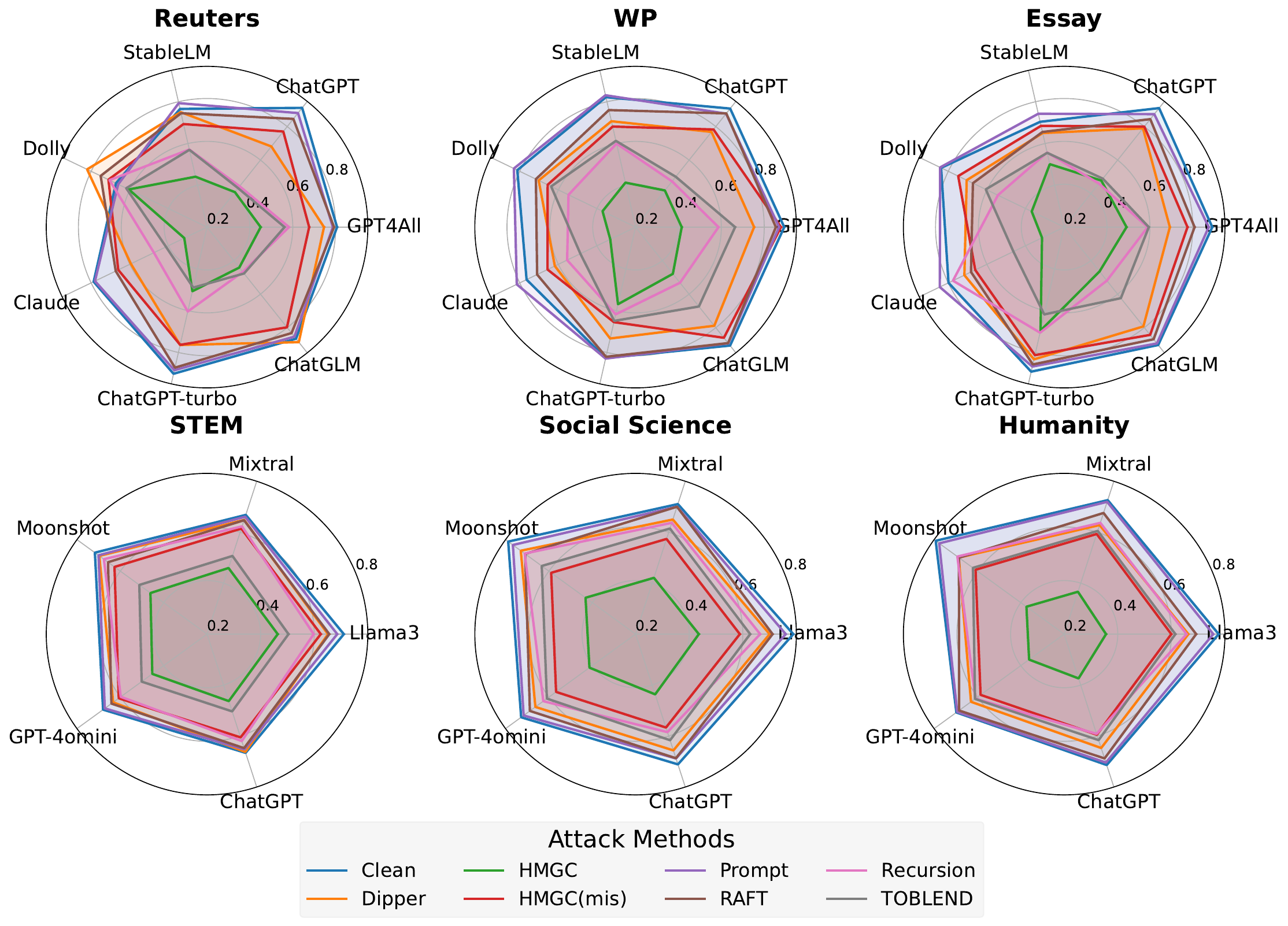} 
    \caption{The AUC performance of evading attacks across LLMs, averaged across detectors.}
    \label{fig:radar} 
\end{figure}

\mypara{Metrics}
For the first task, \method provides a diverse set of metrics for performance evaluation, including accuracy, precision, recall, F1-score, and AUC (Area Under the ROC Curve). 
In the main part of this paper, we use AUC as the evaluation metric, which considers both the True Positive Rate (TPR) and False Positive Rate (FPR) across varying classification thresholds, making it useful for assessing attack performance~\cite{mitchell2023detectgpt}. 
For the second task, we evaluate text fluency using Perplexity (PPL) computed by the \texttt{GPT-2}, which measures how well a language model predicts a given text, with lower PPL indicating higher fluency~\cite{zheng2024cl}.
Semantic similarity is assessed using the cosine similarity (CS) of embeddings from \texttt{all-MiniLM-L6-v2},\footnote{\url{https://huggingface.co/sentence-transformers/all-MiniLM-L6-v2}.} which captures deep semantic relationships, along with ROUGE-L (R-L), which measures lexical overlap based on the longest common subsequence~\cite{lin2004rouge}.
To evaluate text complexity, we use the Flesch Reading Ease (FRE) score, which quantifies readability based on sentence length and word difficulty, with higher scores indicating more accessible text~\cite{farr1951simplification}.
For the final task, we evaluate both runtime and memory consumption on a server equipped with eight NVIDIA L20 GPUs (48GB VRAM each) and dual Intel Xeon Platinum 8369B CPUs (128 cores in total). 

\mypara{Evading Attacks}
\method benchmarks the evading attack that we mentioned in ~\cref{subsec:adv}. 
For Dipper~\cite{krishna2024paraphrasing}, we use their officially released model.\footnote{\url{https://huggingface.co/kalpeshk2011/dipper-paraphraser-xxl}.}
We employ Dipper~\cite{krishna2024paraphrasing} as the base model for Recursion~\cite{sadasivan2023can}, setting the recursion depth to 5, following the default configuration provided in the original code.
For Prompt~\cite{xu2023llm}, we retain the same prompt format and content as the paper and use~\texttt{GPT-4o-mini}~\footnote{\url{https://chatgpt.com/?model=gpt-4o-mini}.} as paraphrase LLM. 
For RAFT~\cite{wang2024raft}, due to the huge token consumption, we opt for the~\texttt{GPT-4o-mini} instead of the \texttt{GPT-3.5-Turbo} model used in the paper, considering the economic cost. 
Apart from this model substitution, all other experimental settings remain consistent with the original paper.
For HMGC~\cite{zhou2024humanizing}, we use the~\texttt{roberta-base}~\footnote{\url{https://huggingface.co/FacebookAI/roberta-base}.} as the surrogate detection model. 
For TOBLEND~\cite{huang2024toblendtokenlevelblendingensemble}, we use the default settings, which are called classic
LLMs set that includes \texttt{GPT-2-xl-1.5B} ~\cite{radford2019language}, \texttt{OPT-2.7B} ~\cite{zhang2022opt}, \texttt{GPT-Neo-2.7B} ~\cite{black2021gpt}, and \texttt{GPT-J-6B}~\cite{wang2021gpt}.
The HMGC framework requires collecting a dedicated dataset to train the surrogate detection model.
The quality of this dataset significantly impacts the detection performance of the surrogate model, thereby affecting the overall effectiveness of the attack.
However, in real-world scenarios, when a user needs to rewrite a sentence, it is impractical to collect a large amount of similar data to train a detector. 
To better evaluate HMGC under more realistic conditions, we propose two experimental settings:
(1) Standard HMGC: The surrogate detection model is trained using the training set corresponding to the test samples. 
(2) Mismatched HMGC: The surrogate detection model is trained on a dataset different from the test set, simulating a real-world scenario where users may not have access to a perfectly matched training dataset. 
Specifically, when evaluating the performance on MGTBench, we train the surrogate model using data from MGT-Academic, and vice versa. 
This setting better reflects practical usage, as users often need to rely on pre-existing datasets that may not perfectly align with their target field. 

\begin{figure}
    \centering
    \includegraphics[width=\columnwidth]{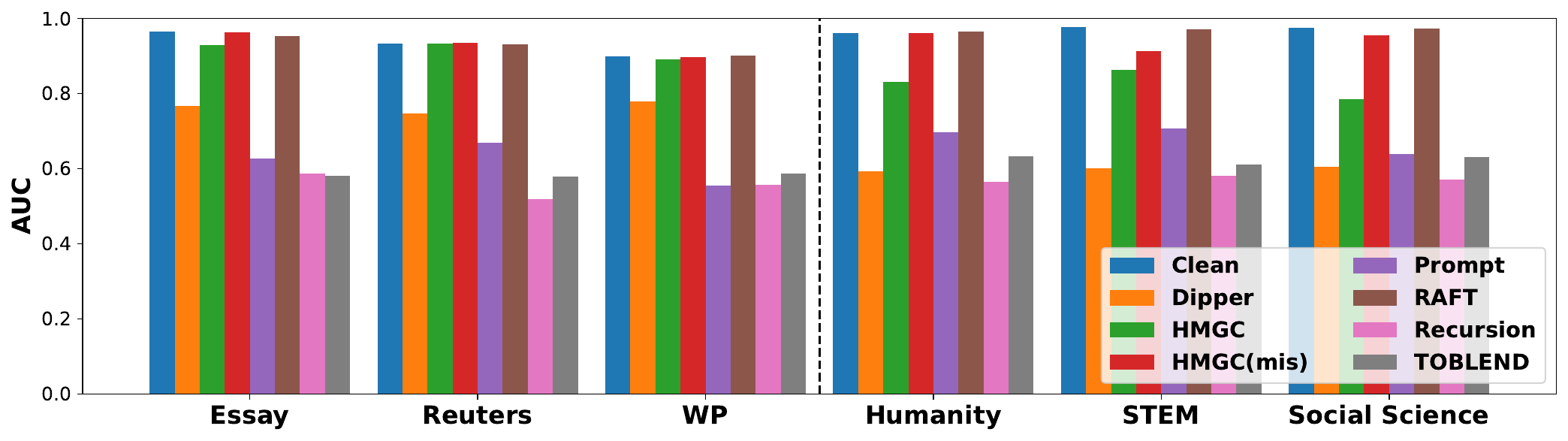} 
    \caption{The AUC performance of Task attribution task for evading attacks against LM-D detector, averaged across LLMs.}
    \label{fig:text bar} 
\end{figure}

\begin{figure}
    \centering
    \includegraphics[width=\columnwidth]{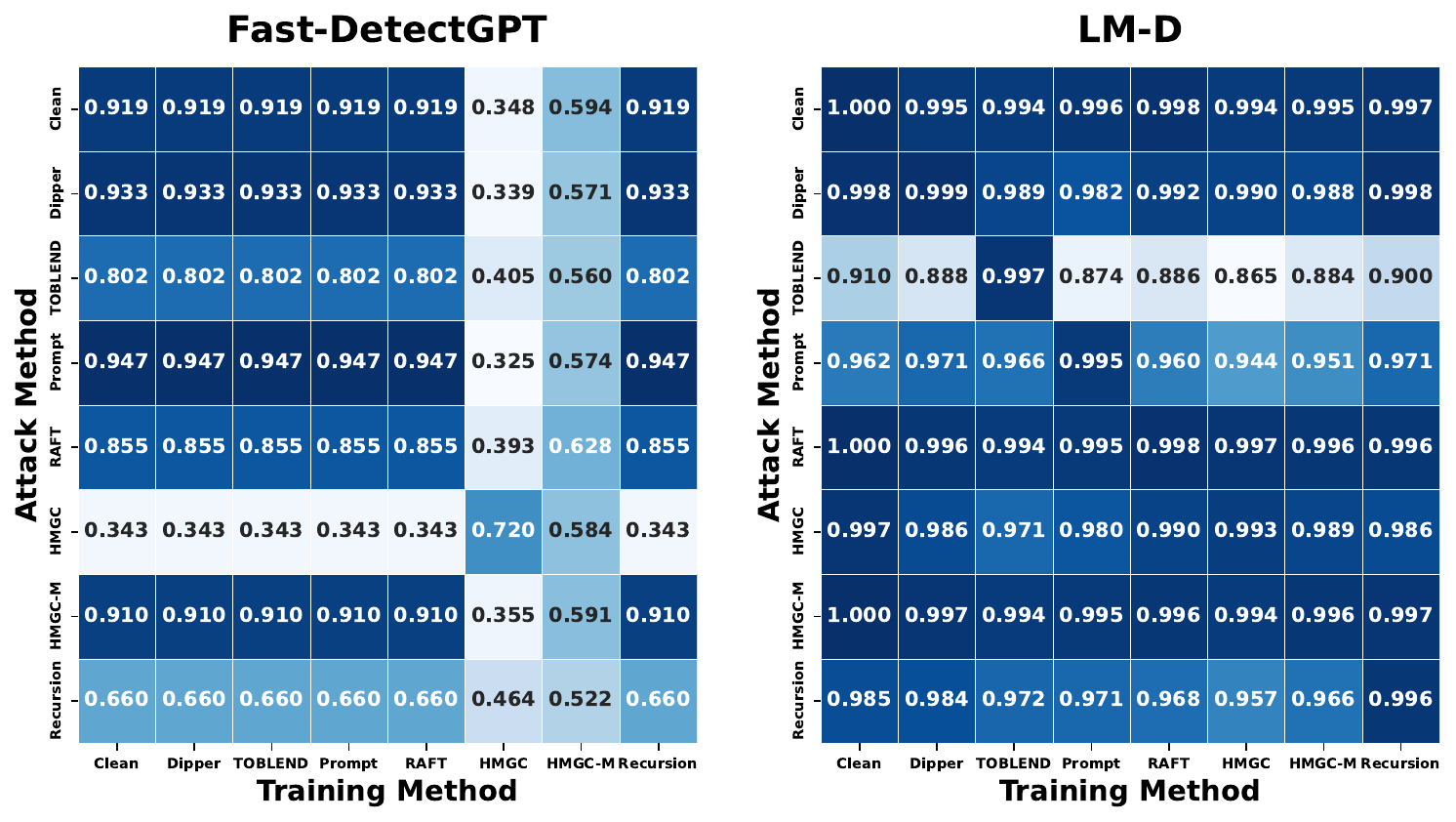} 
    \caption{The AUC value that evaluates attack effectiveness in a real-world scenario.}
    \label{fig:heat} 
\end{figure}

\begin{table*}
  \caption{Text Quality Assessment results: higher CS and R-L values are better, while PPL and FRE are better when closer to the original. Values are averaged across different LLMs.}
  \label{tab:Text Quality}
  \resizebox{\textwidth}{!}{
  \begin{tabular}{c|c|c|c|c|c|c|c|c|c|c|c|c}
        \toprule
        &\multicolumn{4}{c|}{Essay}&\multicolumn{4}{c|}{WP}&\multicolumn{4}{c}{Reuters} \\
        \cmidrule{2-13}
    \multirow{-2.7}{*}{\textbf{\makecell{Dataset $\rightarrow$\\Metric $\rightarrow$\\Method $\downarrow$}}}  &\makecell{PPL \\(Org: 16.778)}&CS &R-L &\makecell{FRE\\(Org: 41.152)}  & \makecell{PPL\\(Org: 19.028)}&CS &R-L &\makecell{FRE\\(Org: 74.965)} &\makecell{PPL\\(Org: 17.727)}&CS &R-L &\makecell{FRE\\(Org: 48.884)} \\
        \cmidrule{1-13}
        Dipper & 33.744 (+16.966) &0.872  &0.379  &50.305 (+9.153) & 23.866 (+4.838) &0.856 &0.399  &81.346 (+6.381)  & 23.687 (+5.96) &0.877 &0.352  &51.839 (+2.955) \\
        
        TOBLEND & 36.113 (+19.335) &0.658  &0.263 &53.654 (+12.502)  & 32.408 (+13.380) &0.569 &0.264  &82.245 (+7.280)  & 41.442 (+23.715) &0.727 &0.281  &57.655 (+8.771)\\
        
        Recursion & 58.165 (+41.387) &0.522  &0.140  &62.881 (+21.729)  & 64.941 (+45.913) &0.445 &0.107  &88.713 (+13.748)  & 96.597 (+78.870) &0.416 &0.088  &67.913 (+19.029)  \\
        
        Prompt & \textbf{16.568 (-0.210)} &0.880  &0.482  &28.238 (-12.914)& \textbf{19.816 (+0.788)} &0.769 &0.465&69.839 (-5.126)   & \textbf{19.660 (+1.933)} &0.895 &0.560  &34.441 (-14.443)  \\
        
        HMGC & 62.539 (+45.761) &0.868  &0.745 &39.660 (-1.492)  & 64.047 (+45.019) & 0.836 &0.746  & 76.939 (+1.974)  & 92.114 (+74.387) &0.833 & 0.691  &49.904 (-1.020)  \\

        HMGC (Mis) & 22.792 (+6.014) &\textbf{0.992}  &\textbf{0.992}  &\textbf{41.221 (+0.069)}  & 21.707 (+2.679) &\textbf{0.991} &\textbf{0.991}  &77.582 (+2.617)  & 20.150 (+2.423) &\textbf{0.990} &\textbf{0.991}  &\textbf{48.765 (-0.119)}  \\
        
        RAFT & 27.472 (+7.694) &0.979 &0.951  &40.722 (-0.430)  & 28.806 (+9.778) &0.982 &0.955 &\textbf{77.108 (+2.143)}  & 27.504 (+9.777) &0.985 &0.962  &48.454 (-0.430)  \\
        
        \midrule
\multirow{-0.7}{*}{\textbf{\makecell{Dataset $\rightarrow$\\Metric $\rightarrow$\\Method $\downarrow$}}}&\multicolumn{4}{c|}{STEM}&\multicolumn{4}{c|}{Social Science}&\multicolumn{4}{c}{Humanity} \\
        \cmidrule{2-13}
        &\makecell{PPL\\(Org: 31.221)} &CS &R-L &\makecell{FRE\\(Org: 30.051)} &\makecell{PPL\\(Org: 25.332)} &CS &R-L &\makecell{FRE\\(Org: 30.481)}&\makecell{PPL\\(Org: 29.343)} &CS &R-L &\makecell{FRE\\(Org: 41.311)} \\
        \cmidrule{1-13}
        
        Dipper & 28.860 (-2.361) &0.729  &0.358  &32.203 (+2.152)   & 30.135 (+4.803) &0.789 &0.376  &42.902 (+12.421)  & 40.006 (+10.663) &0.789 &0.373  &51.157 (+9.846)  \\
        
        TOBLEND & 48.396 (+17.175)  &0.648  &0.287  &54.931 (+24.880)  & 38.846 (+13.514) &0.683 &0.303  &48.496 (+18.015)  & 43.934 (+14.591) &0.655 &0.286  &59.322 (+18.011) \\
        
        Recursion & 125.989 (+94.768) &0.533  &0.238  &46.779 (+16.728)  & 54.909 (+29.577) &0.604 &0.269  &51.301 (+20.820)  & 63.353 (+34.010) &0.611 &0.274  &58.291 (+16.980)  \\
        
        Prompt & \textbf{29.538 (-1.683)} &0.858  &0.596 &\textbf{29.164 (-0.887)}  & \textbf{26.570 (+1.240)} &0.904 &0.603  &27.828 (-2.653)  & \textbf{30.249 (+0.906)} &0.919 &0.613  &38.311 (-3.000)  \\
        
        HMGC & 86.140 (+54.919) &0.814  &0.785  &34.217 (+4.166)  & 99.185 (+67.964) &0.805 &0.756  &\textbf{30.194 (+0.143)}  & 99.280(+69.937) &0.800 &0.758  &47.978 (+6.667) \\

        HMGC (Mis) & 64.250 (+33.029) &0.766 &0.712  &33.502 (+3.451)  & 68.695 (+43.363) &0.824 &0.752 &32.562 (+2.081)  & 79.072(+49.729) &0.848 &0.798  &\textbf{40.973 (-0.338)} \\
        
        RAFT & 37.547 (+6.326) &\textbf{0.987}  &\textbf{0.965}  &31.921 (+1.870)  & 37.254 (+11.922) &\textbf{0.992} &\textbf{0.965}  &35.300 (+4.819)  & 42.220 (+12.877) &\textbf{0.984} &\textbf{0.958}  &35.171 (-6.140)  \\
        
        \bottomrule
    \end{tabular}}
\end{table*}

\section{Experiments and Analysis}

\begin{figure}[htbp]
    \centering
    \includegraphics[width=\columnwidth]{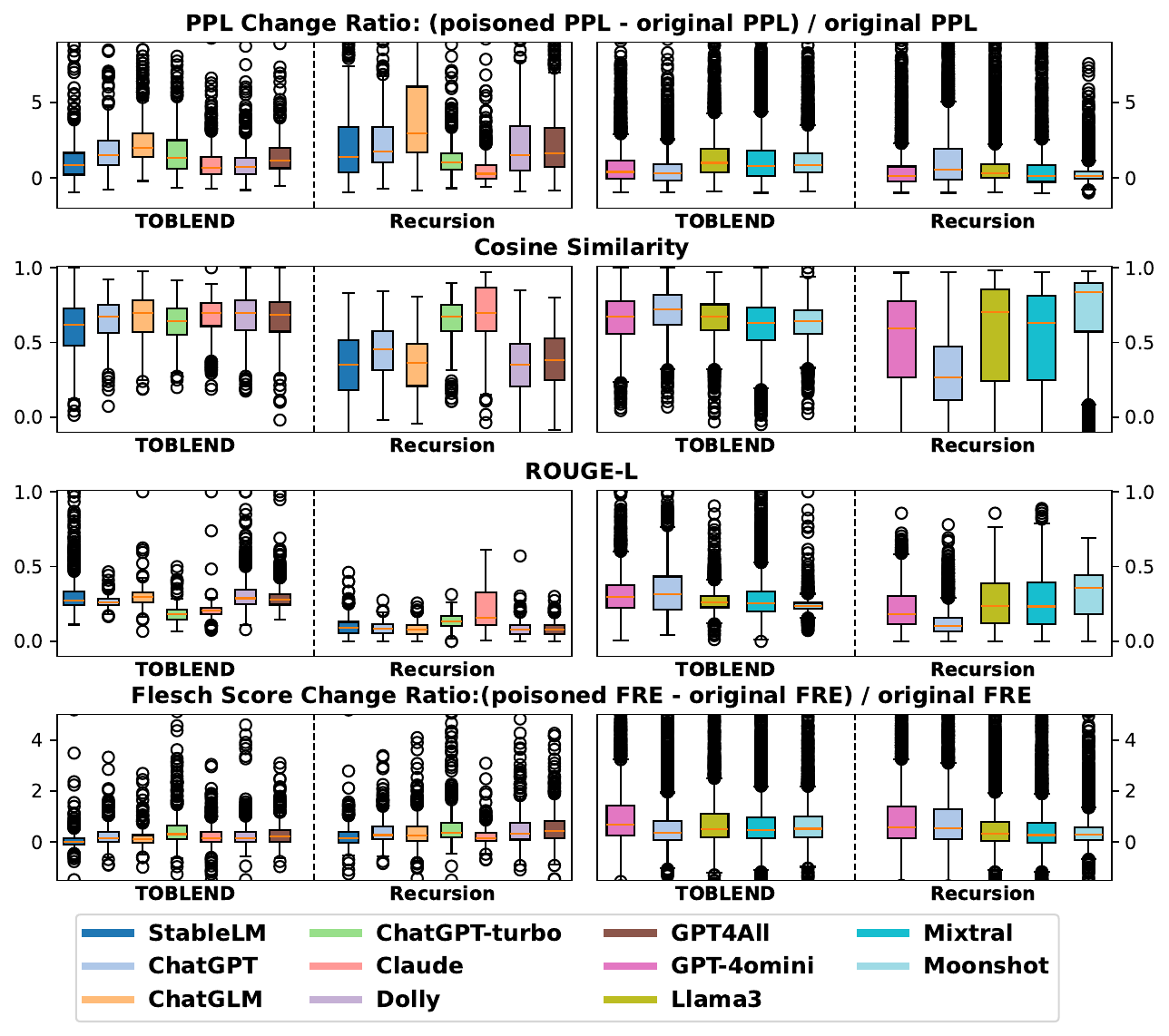} 
    \caption{Text quality comparison of different LLMs: left panel for MGTBench, right for MGT-academic, with values averaged across all fields within each dataset.}
    \label{fig:box} 
\end{figure}

\subsection{Attack Effectiveness Assessment}

In \method, we first evaluate the effectiveness of evading attacks.
The result of the binary classification task (MGT or not)  can be found in~\Cref{tab:big}.
We find that the standard HMGC on different datasets performs well when attacking metric-based detectors. 
For instance, for Log-Likelihood, the AUC drops significantly from 0.913 to 0.185 on Essay and from 0.808 to 0.020 on Reuters.
On the other hand, Prompt and RAFT demonstrate less satisfactory performance across all datasets.
For example, for Log-Likelihood on WP, the AUC of Prompt unexpectedly increases from 0.907 to 0.928,  while RAFT shows only a minimal drop of 0.002 on the STEM dataset, from 0.671 to 0.669.
Another notable phenomenon is that Mismatched HMGC demonstrates some attack effectiveness on the STEM, Social Science, and Humanities datasets. 
However, it shows little to no effectiveness on Essay, WP, and Reuters.
We suspect this is due to the more generalizable features present in MGT-Bench data. 
In Mismatched HMGC, the surrogate detection models for STEM, Social Science, and Humanities are trained on Essay, WP, and Reuters, and vice versa. 
Since the effectiveness of the surrogate model directly impacts the attack performance of HMGC (which also explains why standard HMGC performs well), this suggests that detectors trained on Essay, WP, and Reuters exhibit stronger transferability.
Regarding attacks against model-based detectors, we find that TOBLEND is effective against ChatGPT-Detector across all datasets, even reducing the AUC from 0.727 to 0 on the WP dataset. 
Meanwhile, Recursion is effective against ConDA, and HMGC also shows some effectiveness against ChatGPT-Detector.
We also show the result of the attack effectiveness on different LLMs, as shown in~\Cref{fig:radar}.
We find that on the Reuters dataset, all attacks show limited impact on the MGT generated by Dolly, with some even leading to increased AUC.

For multi-class classification tasks, the goal is to attribute the text to its source model, i.e., identifying which LLM generated it. 
In this experiment, we employ LM-D as the representative detector, as it demonstrated the best performance in our binary classification tasks. 
The experimental results are presented in~\Cref{fig:text bar}, while more detailed results are in~\Cref{sec:task1}. 
Unlike in binary classification, where TOBLEND achieved the most effective attack against LM-D across all datasets, Recursion demonstrates good performance in the multiclass classification task.
Additionally, while Dipper and Prompt showed limited attack effectiveness in binary classification, they perform notably well in the multiclass classification task.

To better simulate the real-world effectiveness of these evading attacks, we explore the scenario where the detectors have been adversarially trained on samples from a known attack and tested on a newly emerging attack (unknown).
To evaluate attack performance in this setting, we select Fast-DetectGPT from metric-based detectors and LM-D from model-based detectors as representative models and use the WP dataset as a case study.
The results are shown in~\Cref{fig:heat}. 
In the Fast-DetectGPT detector, we observe that HMGC consistently achieves the best attack performance, regardless of the training methods.
However, when the training method is HMGC, all attacks (except itself) show low AUC values, indicating that the HMGC training data has significantly compromised the effectiveness of the detector. 
The results of other attack methods on the Fast-DetectGPT detector, along with those from the LM-D detector, indicate that the effectiveness of these attacks does not significantly decline when the detector has been trained with a specific known attack method.
This highlights the independence of these attack methods and demonstrates a certain level of robustness.

In general, we consider HMGC as the most effective evading attack on metric-based detectors since it can decrease the AUC value the most. 
However, its assumption of collecting similar datasets is too strong, limiting its practical applicability.
On the other hand, for model-based detectors, no single evading attack is effective against all detectors.
However, there are some targeted relationships, such as TOBLEND being particularly effective against ChatGPT-Detector, while Recursion is ineffective against the OpenAI-detector, even increasing the probability of detection (AUC increased).

\begin{figure}
    \centering
    \includegraphics[width=\columnwidth]{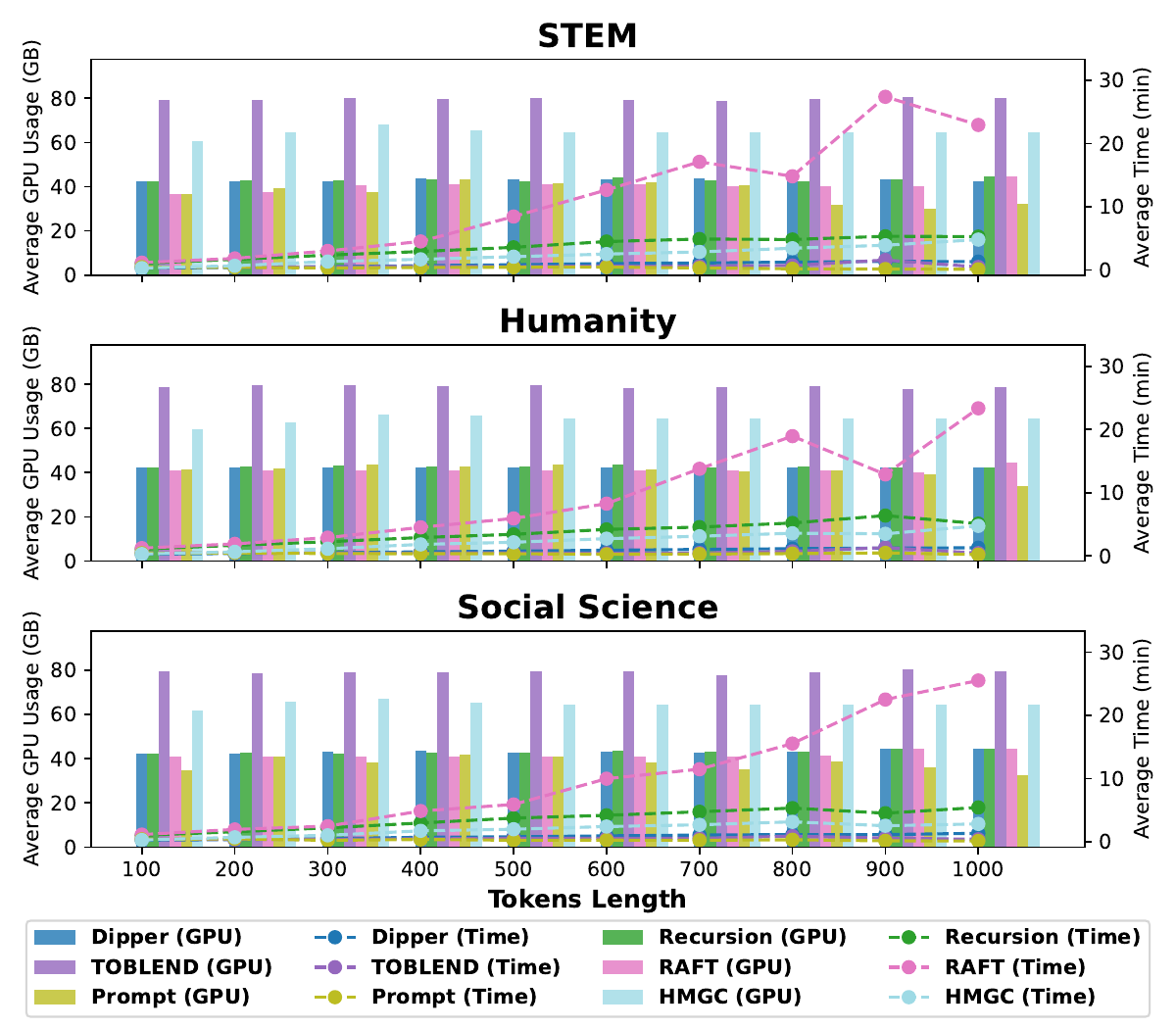} 
    \caption{GPU and time overhead of different attacks, with values averaged across LLMs.}
    \label{fig:usage} 
\end{figure}

\subsection{Text Quality Assessment}

In \method, we also evaluate text quality after different attacks from three aspects: text fluency (measured by PPL), text semantic consistency (measured by Cosine Similarity and ROUGE-L), and text complexity (measured by FRE).
As shown in~\Cref{tab:Text Quality}, we observe that the text quality after the Recursion attack declines noticeably. 
Specifically, text fluency drops significantly across all datasets, with Cosine Similarity falling below 0.65 and ROUGE-L scores consistently below 0.3, indicating a notable loss in semantic consistency, while text complexity also undergoes substantial changes.
On the other hand, the Prompt attack demonstrates a PPL almost identical to the original text in terms of fluency. 
We think this is because the Prompt attack directly generates full sentences using LLMs, which results in more fluency compared to other attack methods that involve partial rewriting.
RAFT shows excellent performance in terms of semantic consistency.
We observe that this is due to RAFT usually rewriting details in sentences, such as prepositions and adverbs, which do not affect the core meaning, thus maintaining good semantic integrity.
In terms of text complexity, except for TOBLEND and Recursion, which generally show poor performance, other methods each excel in different datasets.
We also show the text quality comparison between different LLMs.
Due to the page limit, part of the result can be found in~\Cref{fig:box} and the full chart can be found in ~\Cref{sec:task2} (\Cref{fig:box1,fig:box2}).
In general, we consider RAFT and Prompt to be effective in maintaining text quality, as they perform well in terms of fluency, complexity, and text similarity. On the other hand, TOBLEND and Recursion perform poorly across all three aspects.

\subsection{Attack Execution Overhead Assessment}

We also evaluate the computational overhead of six attacks using data from five LLMs across three datasets of MGT-Academic.
We randomly sample 10 MGTs for each LLM and dataset, with lengths ranging from 100 to 1,000 tokens in steps of 100.
Specifically, for each target length, we select texts that are longer than the target but shorter than the target plus 100, then truncate them to the exact target length. 
This way yields up to 100 MGTs for our overhead evaluation.
In certain LLMs, MGTs in specific fields do not reach 700 to 1,000 tokens, leading to incomplete assessment across the full-length range. The number of different length MGTs for each LLM and field can be found in~\Cref{sec:task3} (\Cref{tab:text_counts}).
We then apply various evading attacks to the sampled MGTs, recording the GPU memory and time usage for each sample.
In this task, we use \texttt{Llama2-7B-chat-hf} instead of \texttt{GPT-4o-mini} for RAFT and Prompt to ensure a fair comparison.
As shown in~\Cref{fig:usage}, the GPU memory overhead does not increase proportionally with token length. 
The minimum memory requirement is close to 40GB, while the highest memory demands are from TOBLEND and HMGC, reaching 80GB and 60GB (with two L20 GPUs), respectively. 
This indicates that all methods have relatively high GPU memory requirements. 
On the other hand, time overhead increases with token length. 
RAFT, with its complex process involving proxy model scoring, word replacement, and candidate word generation, has a significantly higher time overhead than other methods. 
In contrast, methods like Dipper and Prompt, which generate text directly, have simpler steps and thus consume less time. 
Recursion, a five-time loop of Dipper, has a time overhead about five times higher than Dipper, consistent with theoretical expectations.
The detailed result of this assessment can be found in~\Cref{sec:task3} (\Cref{fig:overhead1,fig:overhead2}).
In general, all methods have relatively high GPU memory requirements, while their time overhead varies significantly depending on the method.

\subsection{Analysis}

In this work, we evaluate various evading attacks from three dimensions: attack effectiveness, text quality, and computational cost. 
\textbf{Our experiments reveal that no evading attack excels across all three dimensions.} 
Among these methods, \textbf{Dipper} maintains high text similarity but performs poorly in attack effectiveness.
In some cases, it even increases the AUC on the Binocular detector compared to clean text, while also significantly altering text fluency and complexity. 
\textbf{TOBLEND} effectively evades most detectors, especially ChatGPT-Detector, but struggles against RARAR with AUC increased across all datasets on the ChatGPT-detector. 
Additionally, TOBLEND is computationally expensive and generates text with mediocre fluency. 
\textbf{Recursion} proves effective against most metric-based detectors but fails against all model-based detectors. Additionally, it significantly degrades text quality.
It has the worst semantic similarity among all methods and, in the STEM dataset, produces text with a PPL nearly four times that of the original. 
\textbf{Prompt} preserves good text quality but generally fails to evade detection across different datasets and detectors. 
\textbf{HMGC} achieves strong results against metric-based detectors but relies on an impractical assumption—it requires a surrogate detector trained on a highly similar dataset.
When the dataset differs, as shown in HMGC (Mis), its attack performance drops significantly.
Additionally, its effectiveness against model-based detectors is limited, and it substantially alters text fluency. 
\textbf{RAFT} shows little effectiveness against either metric-based or model-based detectors but performs poorly otherwise. 
However, it maintains good text quality across fluency, similarity, and complexity. 
The main drawback is its extremely high computational cost, making it less practical for large-scale attacks. 

We show the normalized results of various attacks we evaluate across three dimensions in~\Cref{fig:fig1}.
Overall, our results suggest that existing evading attacks struggle to balance attack effectiveness, text quality, and computational efficiency simultaneously.

\section{Discussion}

After evaluating various evading attacks, we have the following insights, which we hope will offer directions for future research.

\begin{tcolorbox}[colframe=gray!50!black, colback=gray!10, coltitle=black, sharp corners=southwest, sharp corners=northeast]
\mypara{\textit{There is a trade-off among attack effectiveness, text quality, and computational cost}}
\end{tcolorbox}

\noindent Our findings suggest an inherent trade-off between attack effectiveness, text quality, and computational cost, forming what appears to be an impossibility triangle in evading attacks against MGT detectors. 
Specifically, when prioritizing attack effectiveness—ensuring the attacked text evades detection—either the quality of the text degrades, making it less natural and coherent, or the computational cost rises significantly due to the need for complex rewriting models. 
Conversely, if we focus on maintaining high text quality and preserving fluency and semantic consistency, the attack becomes less effective, as minor perturbations are often insufficient to bypass detection. 
Lastly, optimizing for low computational cost typically involves lightweight transformations that sacrifice either attack success or text quality. 
We show experiments and analyses in~\Cref{sec:trade_off} to demonstrate this idea.

\begin{tcolorbox}[colframe=gray!50!black, colback=gray!10, coltitle=black, sharp corners=southwest, sharp corners=northeast]
\mypara{\textit{Optimization Insight 1: Quality-Preserving Attack could be a simple but effective plug-in to improve  text quality without compromising attack effectiveness and with minimal computational overhead}}
\end{tcolorbox}

\noindent To improve the trade-off among attack effectiveness, text quality, and computational cost, we propose an insight called \attack (QPA), which can be integrated with existing attack methods to enhance fluency, coherence, and semantic integrity while maintaining competitive attack performance.
Simply put, QPA introduces text quality constraints during adversarial text generation, ensuring that the output remains more fluent, coherent, and semantically intact. 
Here, we use Prompt, RAFT, and TOBLEND as representatives of three distinct types of attacks: paraphrase, perturbation, and data mixing.
For the Prompt attack, we enhance the LLM input with a specific instruction to preserve text quality while generating adversarial content.
For the RAFT attack, we apply quality constraints during word selection, aiming to reduce detection success while maintaining the original PPL, FRE, and high semantic similarity.
For the TOBLEND attack, we rank and select the highest-quality token from multiple LLM outputs instead of choosing randomly, ensuring better text fluency and coherence.
Our detailed settings, experiments, and results in~\Cref{sec:QPA} demonstrate the effectiveness of QPA in improving text quality without compromising attack effectiveness and with minimal computational overhead.

\begin{tcolorbox}[colframe=gray!50!black, colback=gray!10, coltitle=black, sharp corners=southwest, sharp corners=northeast]
\textbf{\textit{Optimization Insight 2: Attack Blending is an effective improvement approach that combines the strengths of two or more evading attacks. } }
\end{tcolorbox}

\noindent The core idea of Attack Blending is to combine different attack strategies to optimize the overall effectiveness. 
To achieve this, we first break down the text into smaller units, such as sentences or even smaller phrases, allowing us to analyze the role each segment plays in the overall structure. 
Some segments have a significant impact on semantic similarity, while others contribute more to the detection accuracy. 
Additionally, some parts are more computationally expensive to modify than others. 
By assessing the characteristics of each segment—such as its effect on the semantic similarity and detection rate—we can select the most appropriate attack strategy for each. 
This insight allows us to prioritize modifying those parts that have a high influence on detection while minimizing changes to parts that have a smaller impact on semantic quality to improve the trade-off among three dimensions.
We show experiments in~\Cref{sec:ab} to illustrate this insight.

\section{Conclusion}

In this work, we introduce \method, the first benchmark for comprehensively evaluating evading attacks against MGT detectors. 
Our benchmark assesses attacks from three key dimensions: evading effectiveness, text quality, and computational overhead. 
Our evaluation of 6 attacks against 13 MGT detectors across 6 datasets generated by 11 LLMs reveals that no existing attack excels in attack effectiveness, text quality, and computational cost.
We further discuss the strengths and limitations of various attacks and identify a trade-off among three dimensions.
Furthermore, we propose two optimization insights and validate their effectiveness through preliminary analysis and experiments.
We hope our benchmark and insights can offer directions for future work.

\bibliographystyle{plain}
\bibliography{arxiv}

\appendix  

\begin{table*}[htbp]
  \caption{Number of MGTs for Different Token Lengths Across Datasets}
  \label{tab:text_counts}
  \resizebox{\textwidth}{!}{
  \begin{tabular}{c|c|c|c|c|c|c|c|c|c|c|c|c|c|c|c}
        \toprule
        \multirow{-0.7}{*}{\textbf{\makecell{Dataset $\rightarrow$\\ token length $\downarrow$}}} & \multicolumn{3}{c|}{Mixtral} & \multicolumn{3}{c|}{Llama3} & \multicolumn{3}{c|}{Moonshot} & \multicolumn{3}{c|}{gpt35} & \multicolumn{3}{c}{GPT-4omini}\\
        \cmidrule(lr){2-4} \cmidrule(lr){5-7} \cmidrule(lr){8-10} \cmidrule(lr){11-13} \cmidrule(lr){14-16}
        &\makecell {STEM} &{Humanities} & {Social Sciences} &\makecell{STEM} &{Humanities} &{Social Sciences} &\makecell {STEM} & {Humanities} & {Social Sciences} &\makecell {STEM} & {Humanities} & {Social Sciences} &\makecell {STEM} & {Humanities} & {Social Sciences} \\
        \midrule
        100 & 10 & 10 & 10 & 10 & 10 & 10 & 10 & 6 & 10 & 10 & 10 & 10 & 10 & 10 & 10 \\
        200 & 10 & 10 & 10 & 10 & 10 & 10 & 10 & 10 & 10 & 10 & 10 & 10 & 10 & 10 & 10 \\
        300 & 10 & 10 & 10 & 10 & 10 & 10 & 10 & 10 & 10 & 10 & 10 & 10 & 10 & 10 & 10 \\
        400 & 10 & 10 & 10 & 10 & 10 & 10 & 10 & 10 & 10 & 10 & 10 & 10 & 10 & 10 & 10 \\
        500 & 10 & 10 & 10 & 10 & 10 & 8 & 10 & 10 & 6 & 10 & 10 & 10 & 10 & 10 & 10 \\
        600 & 10 & 10 & 10 & 3 & 10 & 5 & 9 & 10 & 6 & 10 & 10 & 6 & 10 & 10 & 10 \\
        700 & 10 & 10 & 10 & 0 & 10 & 9 & 3 & 10 & 3 & 10 & 10 & 3 & 10 & 10 & 10 \\
        800 & 0 & 3 & 2 & 0 & 10 & 0 & 0 & 10 & 3 & 5 & 2 & 0 & 10 & 10 & 10 \\
        900 & 0 & 0 & 0 & 0 & 0 & 0 & 0 & 1 & 0 & 1 & 0 & 0 & 10 & 10 & 10 \\
        1000 & 0 & 0 & 0 & 0 & 0 & 0 & 0 & 0 & 0 & 0 & 0 & 0 & 10 & 10 & 10 \\
        \bottomrule
    \end{tabular}}
\end{table*}

\section{Related Work}
\label{sec:Related}

To the best of our knowledge, no study has comprehensively evaluated evading attacks against MGT detectors. 
However, several benchmarks have evaluated detectors in adversarial settings, addressing some aspects of evading attacks. 
MGTBench~\cite{he2024mgtbench} demonstrates the impact of three types of attacks on binary classification (whether the text is MGT), but it does not test how evasion attacks affect other tasks or broader aspects. 
Stumbling Blocks~\cite{wang2024stumbling} provides a more comprehensive analysis of the impact of different attack types on binary classification tasks and includes preliminary tests on text quality changes, such as semantic similarity and fluency. 
However, it does not explore how attacks affect multi-class tasks and a wider range of text quality metrics (such as text complexity), nor does it account for the computational costs of these attacks. 
This highlights the different focus between benchmarks centered on detector evaluation and those focused on evading attacks. 
Finally, DetectRL~\cite{wu2024detectrl} involves the generalizability of detectors, showcasing how adversarial training with one evasion attack improves resistance to others. 
However, it still lacks evaluations of text quality, computational costs, and multi-task performance.

Detector-centered benchmarks often focus on the impact of evading attacks on a single aspect, such as binary classification performance, typically in idealized settings without considering resource costs or text variations. 
In contrast, evading attack-centered benchmarks provide a more comprehensive evaluation, assessing how evading attacks affect multiple aspects, including text quality, task performance, and computational cost. 
This broader perspective helps to understand the feasibility of these attacks and supports the development of more holistic optimization strategies.

\begin{figure}
    \centering
    \includegraphics[width=\columnwidth]{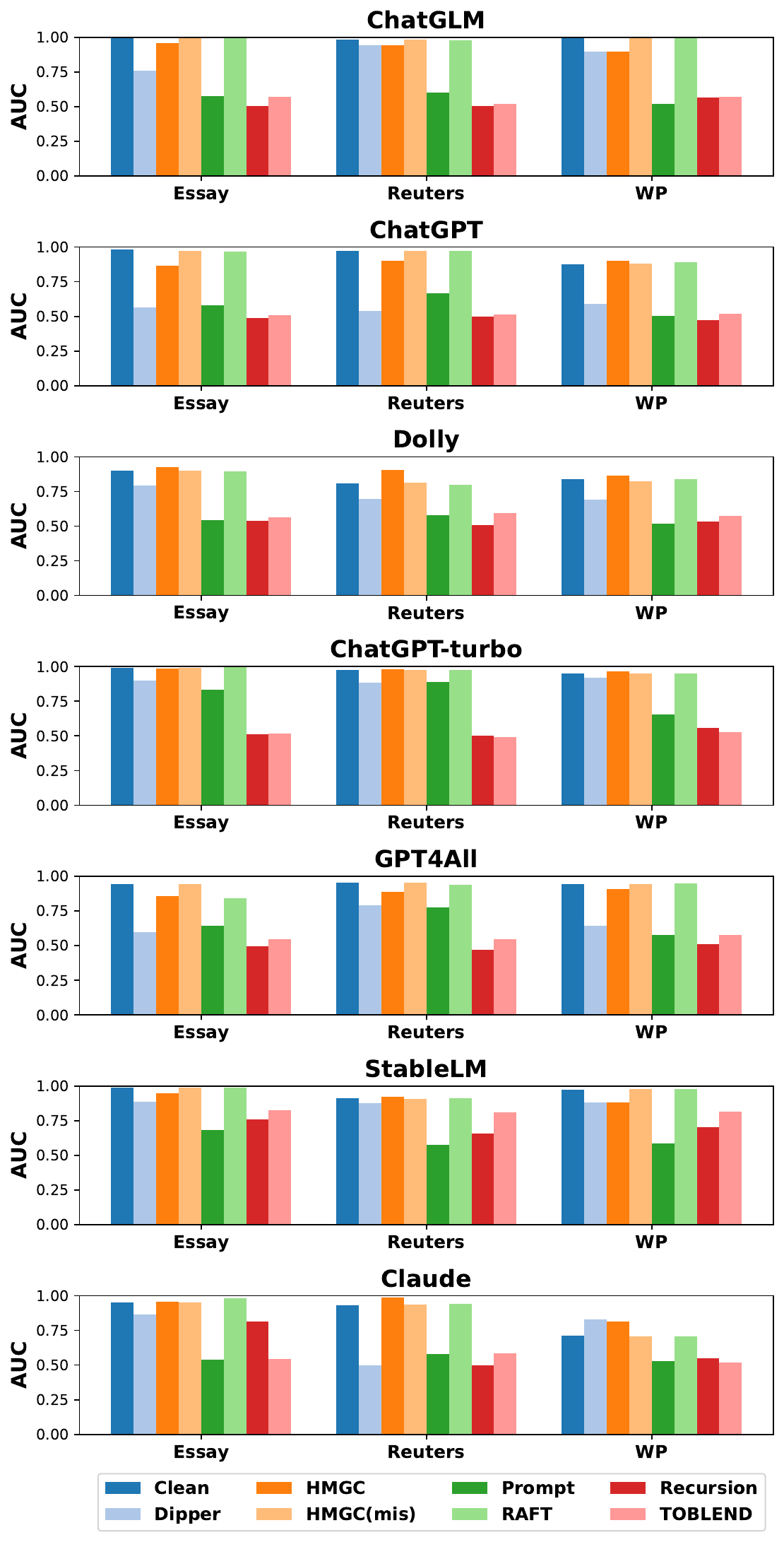} 
    \caption{The AUC performance of Task attribution task for evading attacks against LM-D detector on MGTBench data. }
    \label{fig:task1_1} 
\end{figure}

\begin{figure}
    \centering
    \includegraphics[width=\columnwidth]{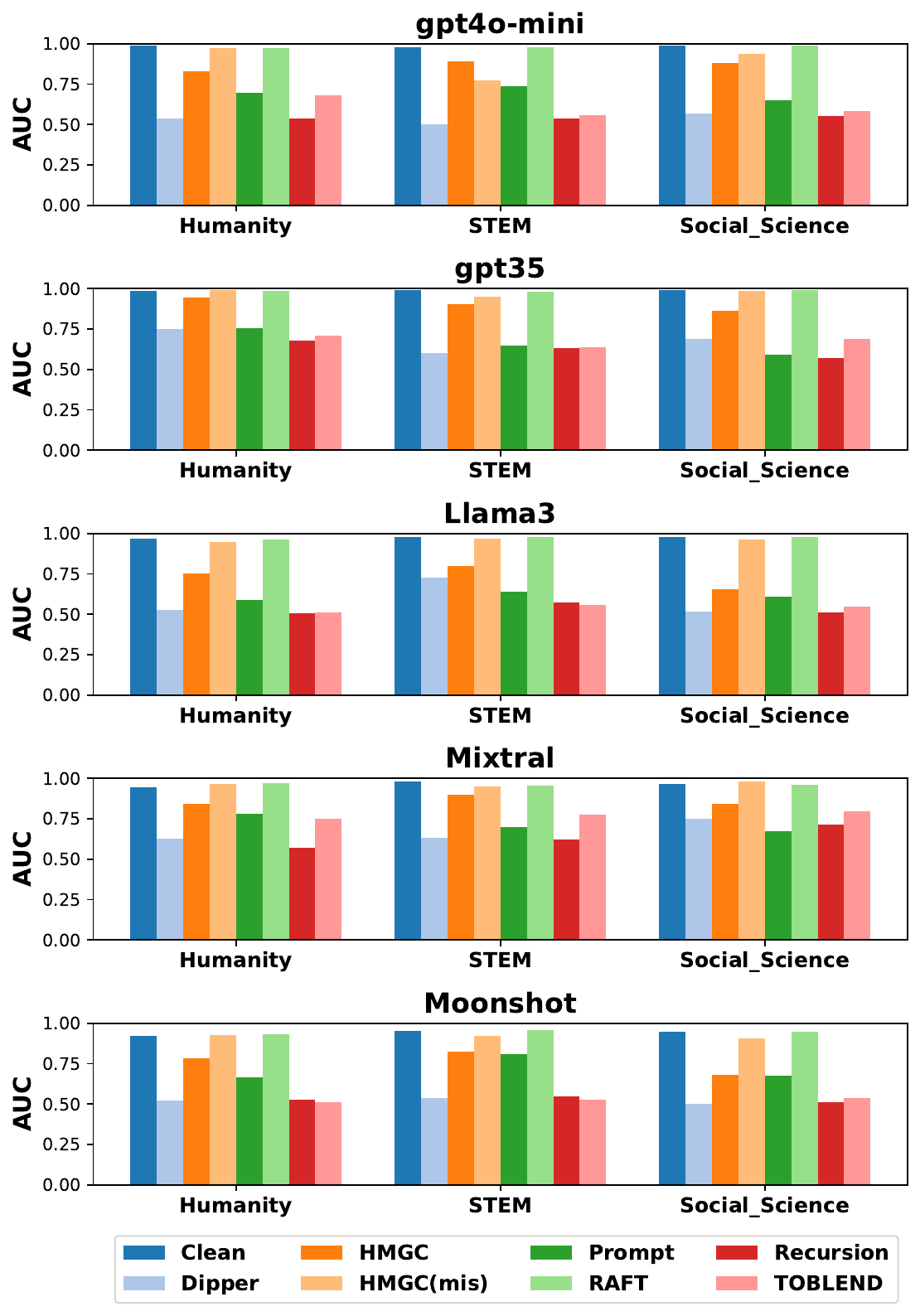} 
    \caption{The AUC performance of Task attribution task for evading attacks against LM-D detector on MGT-Academic data.}
    \label{fig:task1_2} 
\end{figure}

\begin{table*}
  \caption{The Text Quality of QPA integrates three representative attacks, with values averaged across LLMs. Higher CS and R-L values are better, while PPL and FRE are better when closer to the original.}
  \label{tab:table3}
  \resizebox{\textwidth}{!}{
  \begin{tabular}{c|c|c|c|c|c|c|c|c|c|c|c|c}
        \toprule
\multirow{-0.7}{*}{\textbf{\makecell{Dataset $\rightarrow$\\Metric $\rightarrow$\\Method $\downarrow$}}}&\multicolumn{4}{c|}{STEM}&\multicolumn{4}{c|}{Social Science}&\multicolumn{4}{c}{Humanity} \\
        \cmidrule{2-13}
        &\makecell{PPL\\(Org: 31.221)} &CS. &R-L &\makecell{FRE\\(Org: 30.051)} &\makecell{PPL\\(Org: 25.332)} &CS. &R-L &\makecell{FRE\\(Org: 30.481)}&\makecell{PPL\\(Org: 29.343)} &CS. &R-L &\makecell{FRE\\(Org: 41.311)} \\
        \cmidrule{1-13}
        
        Prompt & 29.538 (-1.683) &0.858  &0.596 & 29.164 (-0.887)  & \textbf{26.570 (+1.240)} &0.904 &0.603  &\textbf{27.828 (-2.653)}  & \textbf{30.249 (+0.906)} &0.919 &0.613    &38.311 (-3.000)  \\

        QPA (Prompt) & \textbf{30.991 (-0.230)} &\textbf{0.964}  &\textbf{0.704}  &\textbf{30.121 (+0.070)}  & 29.200 (+3.868) &\textbf{0.957} &\textbf{0.673}  & 34.105 (+3.624)  & 37.564 (+8.221) &\textbf{0.955} &\textbf{0.683}  &\textbf{44.169 (+2.858)}  \\
        
        \cmidrule{1-13}
        RAFT & 37.547 (+6.326) &0.987&0.965  &31.921 (+1.870)  & 37.254 (+11.922) &0.992 &0.965  &35.300 (+4.819)  & \textbf{42.220 (+12.877)} &0.984 &0.958  &35.171 (-6.140)  \\

        QPA (RAFT) & \textbf{34.503 (+3.282)} &\textbf{0.997}  &\textbf{0.983}  &\textbf{30.779 (+0.728)}  & \textbf{34.909 (+9.577)} &\textbf{0.998} &\textbf{0.986}  &\textbf{32.621 (+6.140)}  & 43.334 (+13.991) &\textbf{0.991} &\textbf{0.964}  &\textbf{38.413 (-2.898)}  \\
        
        \cmidrule{1-13}
        TOBLEND & 48.396 (+17.175)  &0.648  &0.287  &54.931 (+24.880)  & 38.846 (+13.514) &0.683 &0.303  &48.496 (+18.015)  & \textbf{43.934 (+14.591)} &0.655 &0.286  &59.322 (+18.011) \\

        QPA (TOBLEND) & \textbf{37.920 (+6.699)}  &\textbf{0.735}  &\textbf{0.408}  &\textbf{48.585 (+18.534)}  & \textbf{29.765 (+4.433)} &\textbf{0.766} &\textbf{0.433}  &\textbf{44.681 (+14.200)}  & 48.428 (+19.085) &\textbf{0.751} &\textbf{0.445}  &\textbf{54.912 (+13.601)} \\
        \bottomrule
    \end{tabular}}
\end{table*}

\section{Datasets}
\label{sec:datasets}

MGTBench dataset contains three fields: Essay, WP, and Reuters. 
The Essay field consists of 1,000 academic essays covering various disciplines at high school and university levels, sourced from IvyPanda\footnote{https://ivypanda.com.}.
The WP field includes 1,000 creative writing pieces derived from the \texttt{r/WritingPrompts} subreddit, where users share writing prompts and corresponding stories.
The Reuters field comprises 1,000 news articles from the Reuters 50-50 authorship identification dataset~\cite{houvardas2006n}, originally written by journalists across various topics. 
For each field, prompts are first constructed using \texttt{ChatGPT-turbo}, and these prompts are then used to generate MGTs from seven different LLMs: \texttt{ChatGPT-turbo}~\cite{ChatGPT}, \texttt{ChatGLM}~\cite{chatglm6b}, \texttt{Dolly}~\cite{Dolly}, \texttt{GPT4All}~\cite{anand2023gpt4all}, \texttt{StableLM}~\cite{StableLM}, \texttt{Claude}~\cite{Claude}, and \texttt{ChatGPT}~\cite{ChatGPT}.
According to~\cite{he2024mgtbench},we randomly
split 80\% of the entries as the training set and the rest as the testing set.

The MGT-Academic dataset consists of three academic parts: STEM, Social Science, and Humanity. 
The STEM part covers physics, math, computer science, biology, electrical engineering, chemistry, medicine, and statistics, with data sourced from Wiki and Arxiv. 
The Social Science part includes disciplines such as education, economy, and management, with data collected from Wiki, Arxiv, and Gutenberg. 
The Humanity part encompasses literature, law, art, history, and philosophy, drawing data from Wiki and Gutenberg. 
All Wiki data is collected from two-level sub-topics within each field.
All Arxiv data is from the abstracts, introductions, and conclusions of papers submitted before 2023 and retains LaTeX formatting to enhance diversity in academic writing. 
All Gutenberg data is randomly sampled paragraphs from books within each field, excluding tables of contents.
Additionally, MGT data is generated using five LLMs: \texttt{Llama-3.1}~\cite{touvron2023llama}, \texttt{Mixtral}~\cite{MistralAI}, \texttt{MoonShot}~\cite{Moonshot}, \texttt{ChatGPT}~\cite{ChatGPT}, and \texttt{GPT-4omini}~\cite{ChatGPT}, which are prompted to act as wiki, paper, or book editors. 
To be consistent with the previous dataset, for each field, we randomly select the same number of training and test sets as in the MGTbench dataset.

\section{Attack Effectiveness Assessment}
\label{sec:task1}

Here, we show the more detailed experimental results of the multiclass classification task on attack effectiveness assessment of various evading attacks.~\Cref{fig:task1_1,fig:task1_2} show the detailed results of different attacks for different LLMs on MGTBench and MGT-Academic data, respectively. We find that the attack effects are similar on texts generated by different LLMs.

\begin{figure}
    \centering
    \includegraphics[width=\columnwidth]{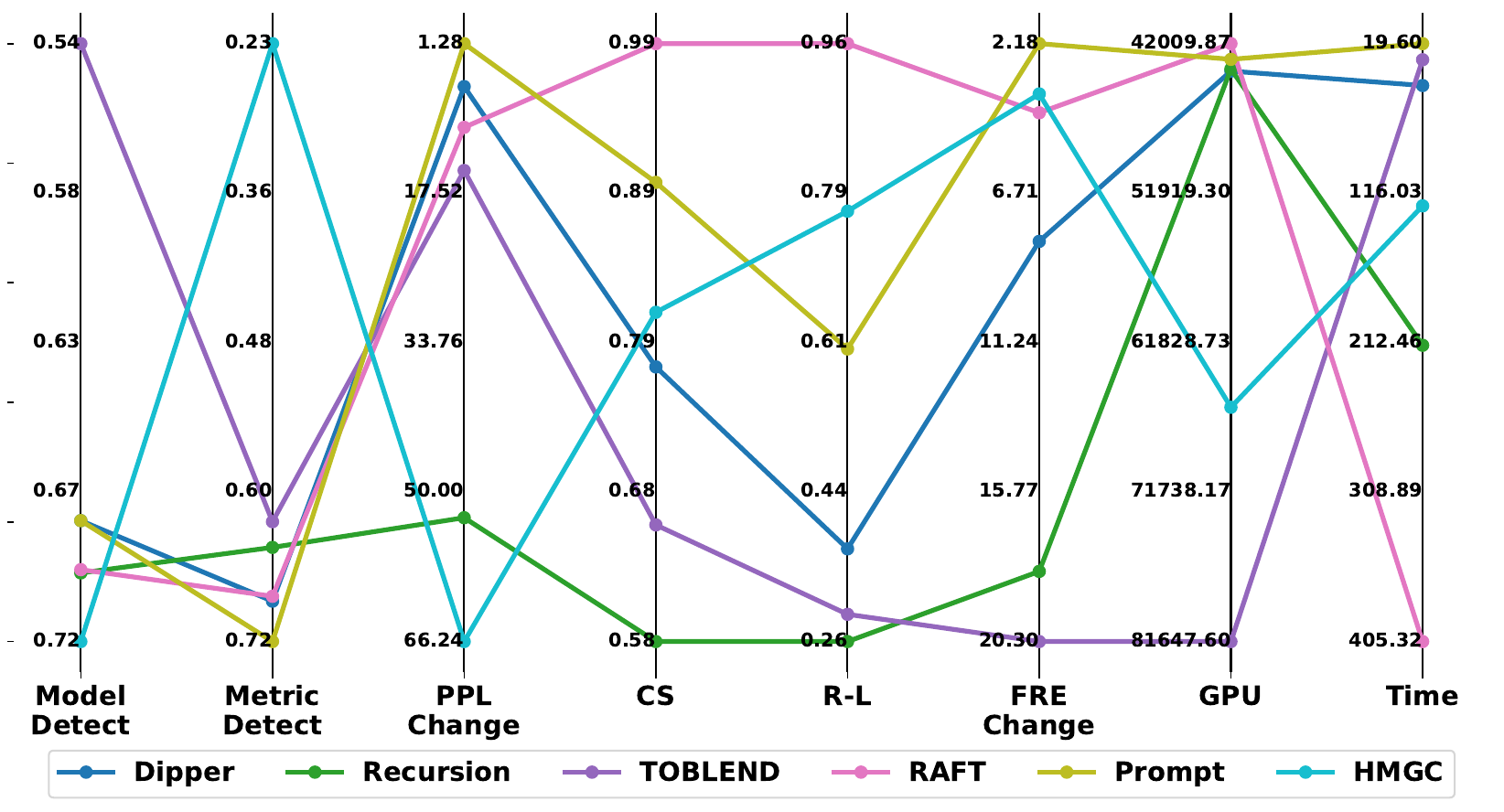} 
    \caption{The trade-offs among the three dimensions are illustrated by using different attack methods. All metrics have been standardized, such that a higher point indicates better performance in the corresponding metric. From this, it is evident that no single method excels across all metrics.}
    \label{fig:trade_off1} 
\end{figure}

\begin{figure}
    \centering
    \includegraphics[width=\columnwidth]{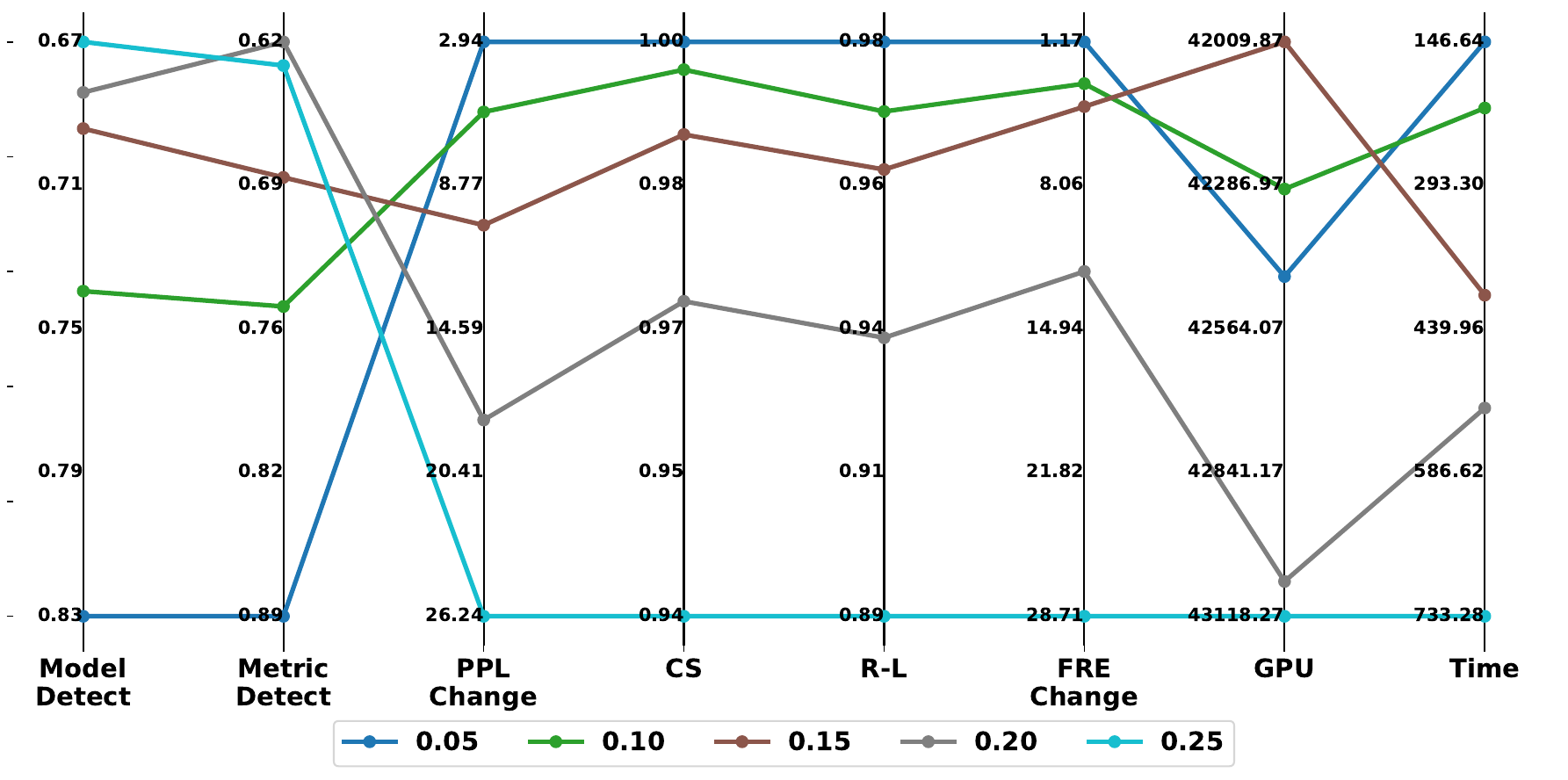} 
    \caption{The trade-offs among the three dimensions are illustrated by using a single attack method. All metrics have been standardized, such that a higher point indicates better performance in the corresponding metric.}
    \label{fig:trade_off2} 
\end{figure}

\section{Test Quality Assessment}
\label{sec:task2}

\begin{figure*}
    \centering
    \includegraphics[width=\textwidth]{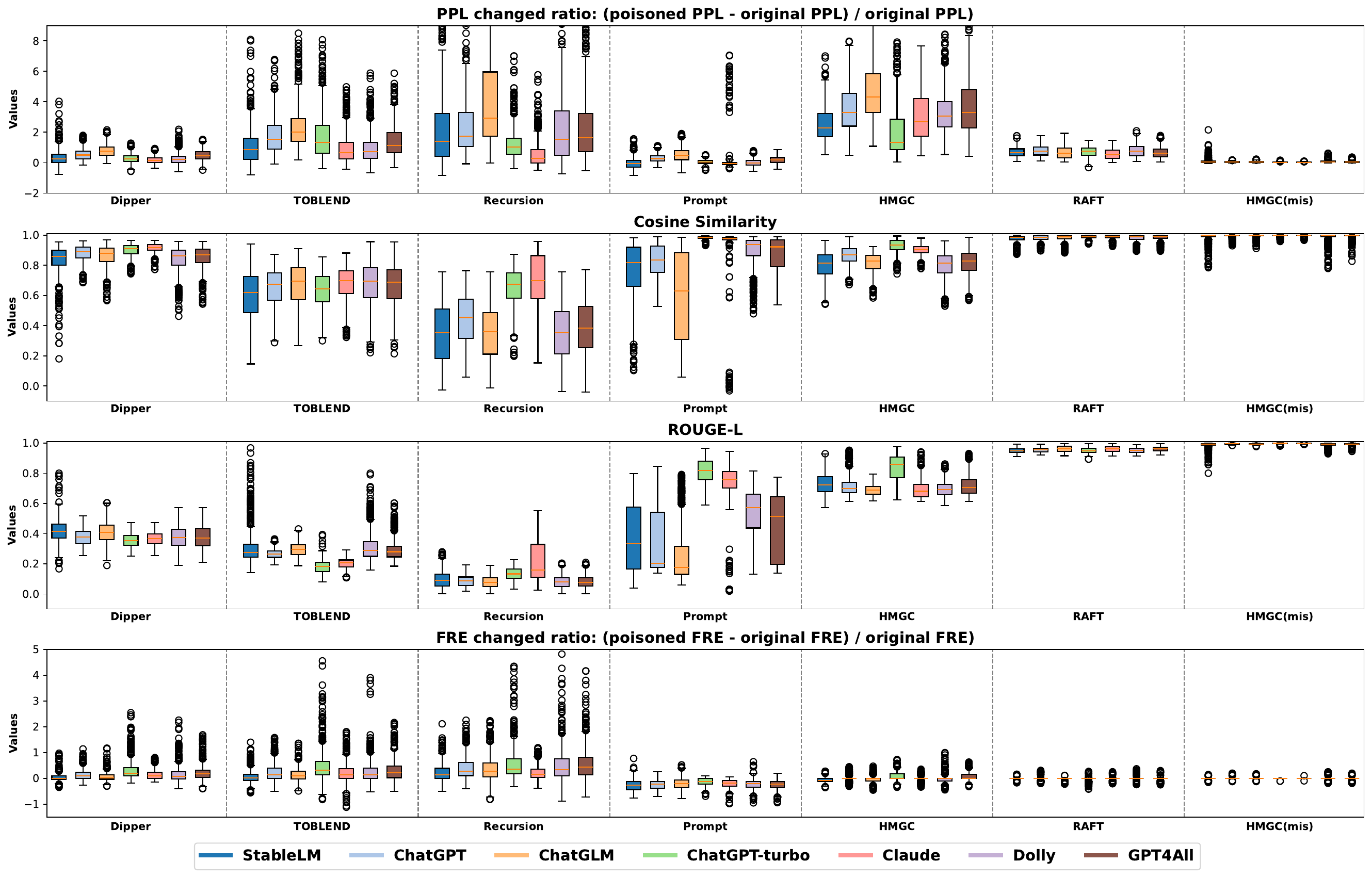} 
    \caption{Text quality comparison of different LLMs in MGTBench data. The values represent the average performance across all fields within each dataset.}
    \label{fig:box1} 
\end{figure*}

\begin{figure*}
    \centering
    \includegraphics[width=\textwidth]{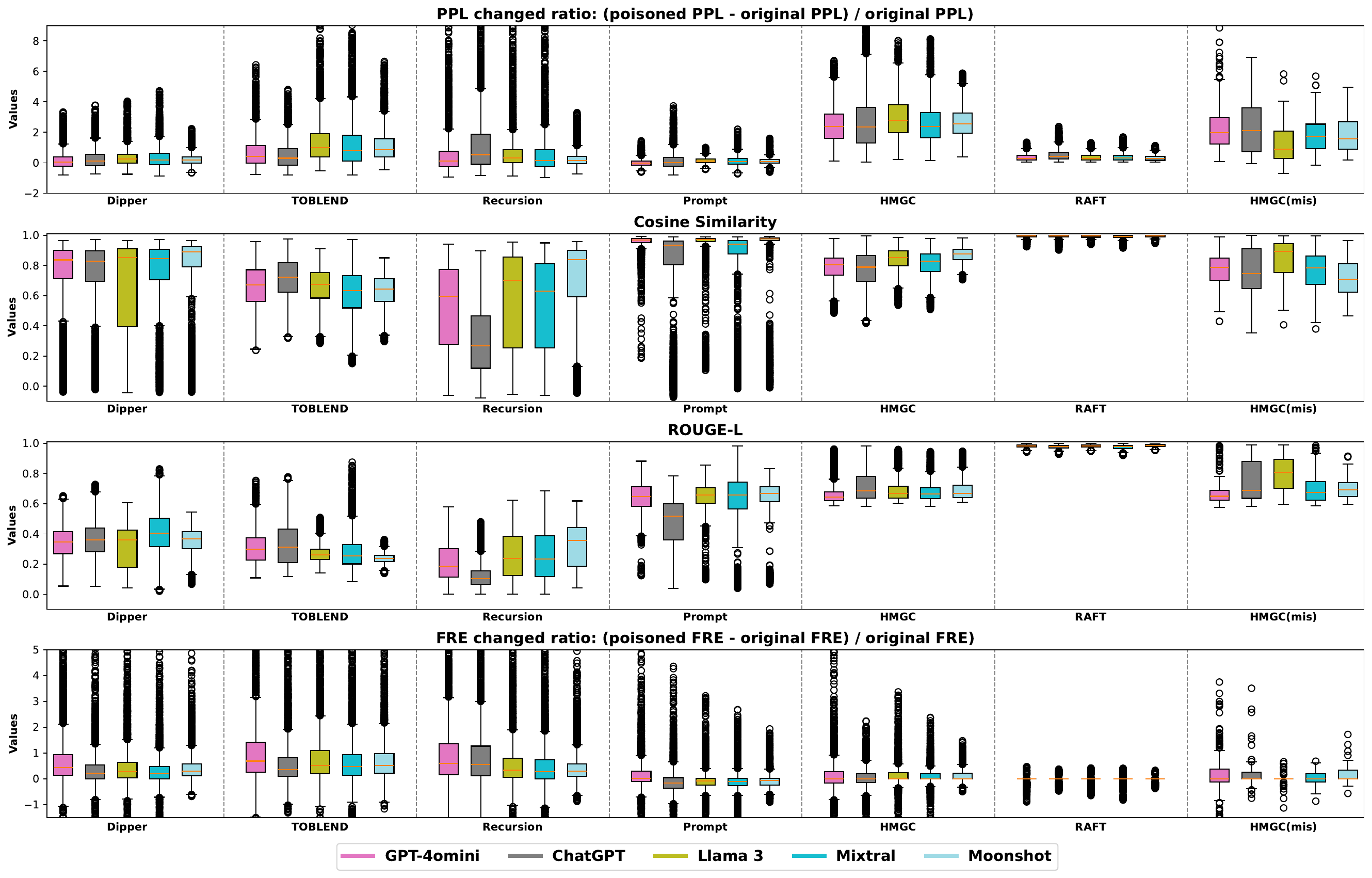} 
    \caption{Text quality comparison of different LLMs in MGT-Academic data. The values represent the average performance across all fields within each dataset.}
    \label{fig:box2} 
\end{figure*}

The~\Cref{fig:box1,fig:box2} show the detailed results of text quality comparison of different LLMs in MGTBench and MGT-Academic data. 
In MGTBench data, the fluctuation of different metrics under text quality in perturbation attacks is minimal. 
In contrast, paraphrase attacks, such as Prompt, exhibit greater volatility across different samples. 
Notably, Prompt performs significantly worse on ChatGLM compared to other LLMs. 
Additionally, Recursion demonstrates poor performance across nearly all LLMs.
Similarly, perturbation attacks also perform well in MGT-Academic data except for the Mismatched HMGC. 
Recursion is still the worst in MGT-Academic data. 

\section{Attack Execution Overhead Assessment}
\label{sec:task3}

The~\Cref{tab:table3} shows the number of MGTs for different token lengths across datasets. 
Among them, the MGT generated by Llama3 in the STEM dataset is the shortest, with only 3 MGTs exceeding 500 tokens. 
Except for GPT-4omini, all other LLMs have some missing MGTs.
The~\Cref{fig:overhead1,fig:overhead2} display the GPU and time consumption of the selected texts during the attack process. 
We observe that the GPU consumption remains stable, with TOBLEND and HMGC utilizing higher GPU resources.
In terms of time, the trends across different datasets and models are quite similar: RAFT exhibits the highest time consumption and the fastest growth, while Prompt, due to requiring fewer operations, consistently shows the lowest time consumption.

\begin{table*}
    \centering
    \caption{Result of Attack Blending. The AUC, PPL change, FRE change, GPU, and Time are lower the better, while CS and R-L are higher the better.}
    \resizebox{\textwidth}{!}{
    \begin{tabular}{c|c|c|c|c|c|c|c|c}
        \toprule
        Attack & Model-based (AUC) & Model-based (AUC)& PPL Change & CS & R-L & FRE Change & GPU (MB) & Time (s) \\
        \midrule
        Dipper & 0.487 & 0.613 & 5.942 & 0.769 & 0.369 & 8.173 & 43832.667 & 46.637 \\
        Attack Blending & 0.498 & 0.618 & 6.862 & 0.888 & 0.832 & 5.426 & 43879.867 & 210.234 \\
        RAFT & 0.541 & 0.651 & 10.375 & 0.988 & 0.963 & 4.276 & 42009.867 & 405.325 \\
        \bottomrule
    \end{tabular}}
    \label{tab:blending} 
\end{table*}

\begin{figure}
    \centering
    \includegraphics[width=\columnwidth]{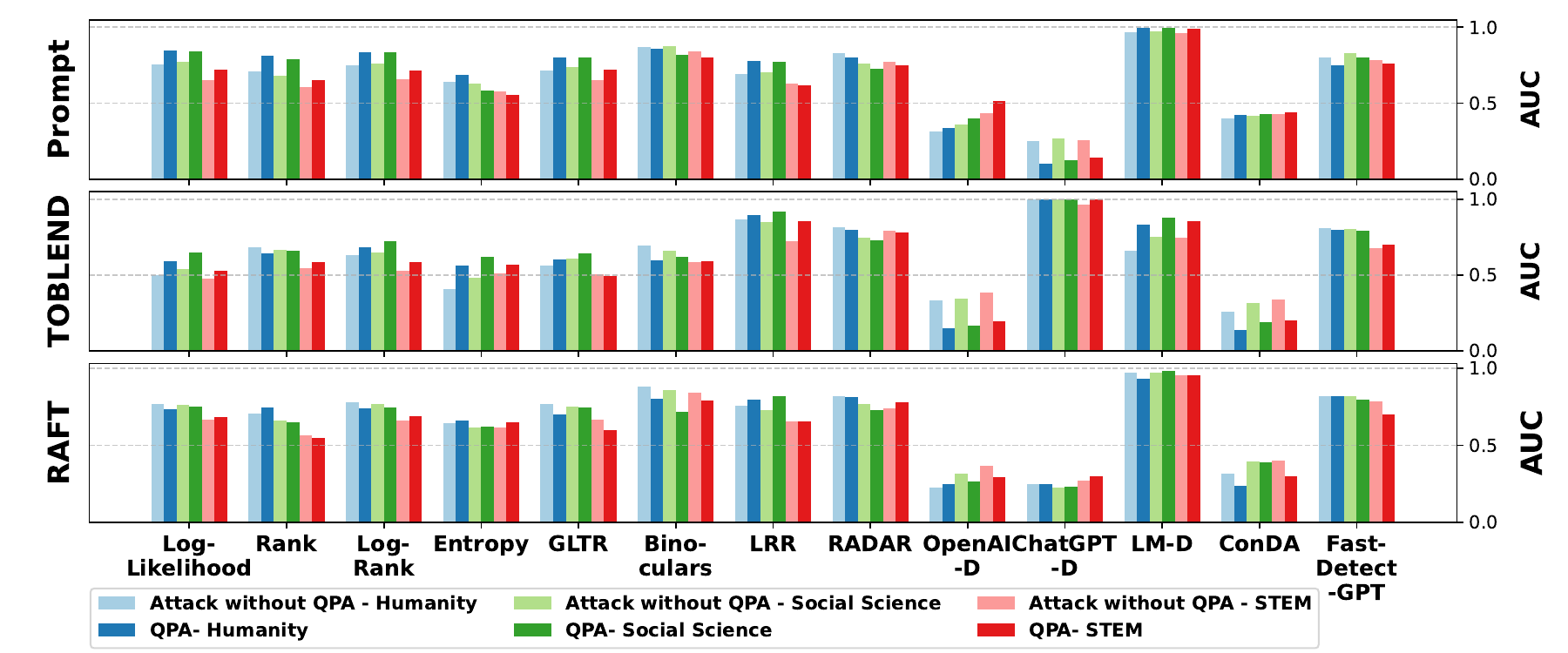} 
    \caption{The Attack Effectiveness of QPA integrates three representative attacks, with values averaged across LLMs. }
    \label{fig:QPA} 
\end{figure}

\begin{figure*}
    \centering
    \includegraphics[width=\textwidth]{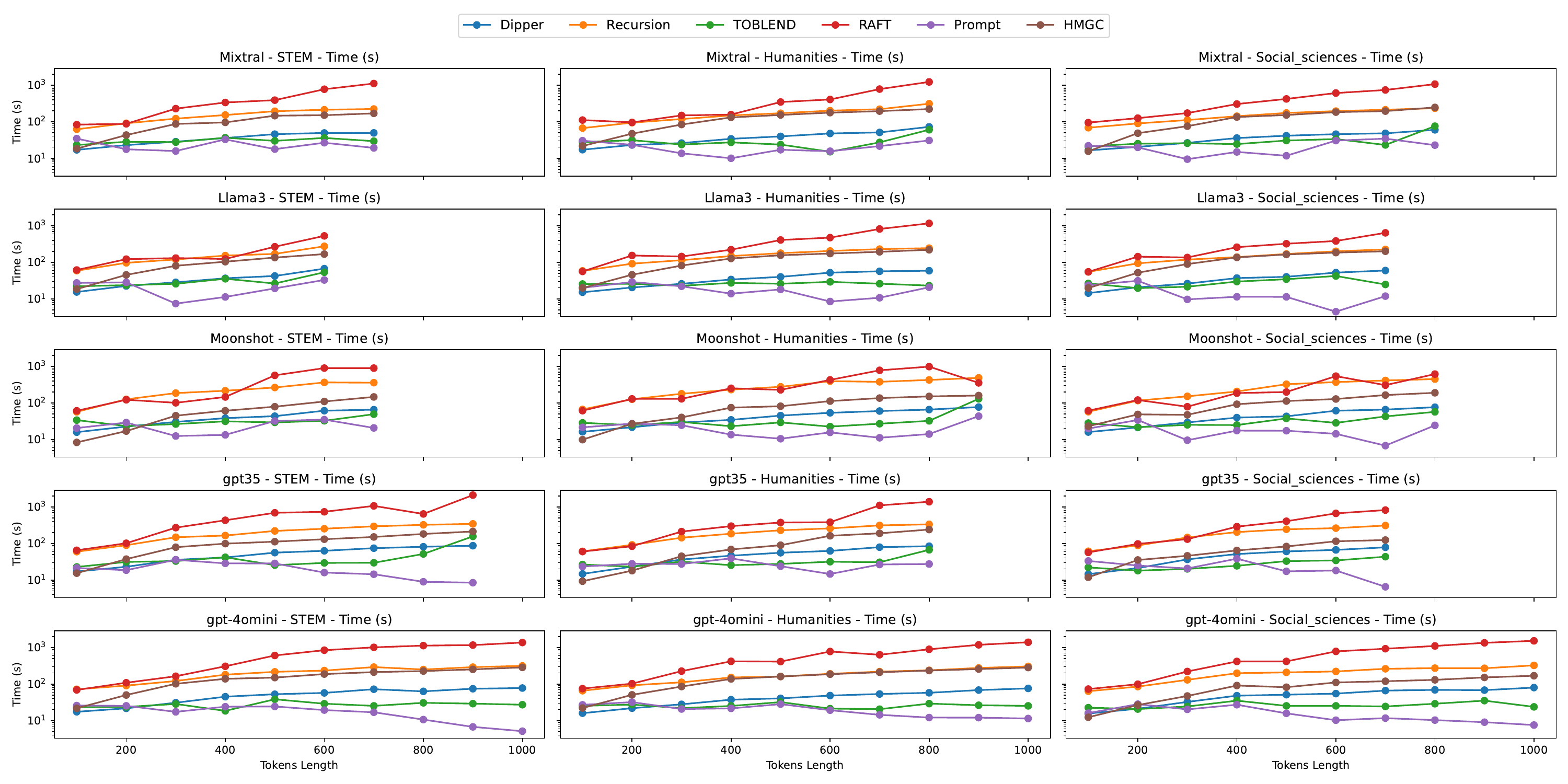} 
    \caption{Time consumption of different attacks in different token lengths.}

    \label{fig:overhead1} 
\end{figure*}

\begin{figure*}
    \centering
    \includegraphics[width=\textwidth]{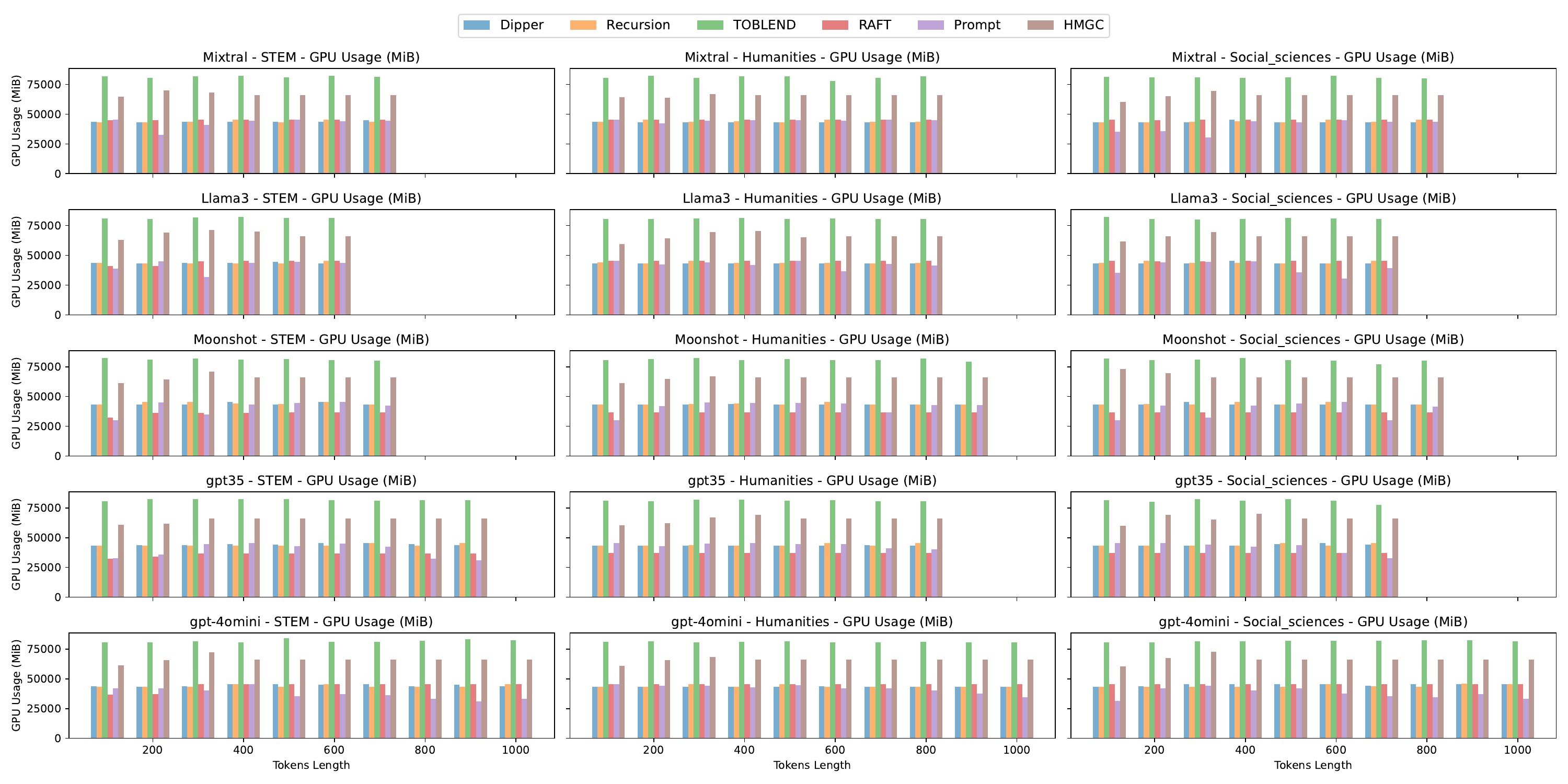} 
    \caption{GPU usage of different attacks in different token lengths.}
    \label{fig:overhead2} 
\end{figure*}

\section{Trade-off among three dimensions}
\label{sec:trade_off}

We present two examples to support the argument that such a trade-off exists: 
Firstly, different attack methods illustrate this phenomenon. 
As shown in~\Cref{fig:trade_off1}, no single method excels across all dimensions. 
Methods that exhibit high attack effectiveness typically incur significant computational overhead and lead to a deterioration in text quality, while methods that prioritize text quality often result in compromised attack effectiveness.
Secondly, we tested this trade-off by adjusting the parameters of a single method to alter the extent of its modifications.
We used RAFT on the WP dataset, where the original proportion of modified words was 0.15. 
In our experiment, we varied this proportion between 0.05 and 0.25 to observe the impact on attack effectiveness, text quality, and computational cost.
The result is shown in~\Cref{fig:trade_off2}.
We find that when the parameter is set to 0.05, the fewest words are modified, allowing the text quality to be well preserved, and the time consumption for the attack is also relatively low. 
However, the attack effectiveness is the weakest. 
As the parameter increases, the attack effectiveness gradually improves, but at the cost of a decline in text quality and an increase in the time required for the attack.

These two examples demonstrate the existence of this trade-off, where optimizing for one aspect—whether it’s attack effectiveness, text quality, or computational cost—inevitably leads to compromises in the other one or two. 
The first example, which compares different attack methods, shows that no single method can achieve optimal performance across all dimensions simultaneously.
Similarly, the second example, where we adjust the modification parameters in RAFT, further highlights the balance between attack effectiveness, text quality, and time consumption. 

\section{Quality-Preserving Attack}
\label{sec:QPA}

We use preliminary experiments to prove the effectiveness of this insight.
For the Prompt attack, we add a specific instruction to the input of the model, requesting it to maintain the quality of the output text while generating adversarial content. 
Specifically, we include the instructions as below:
\begin{tcolorbox}[colframe=gray!50!black, colback=gray!10, coltitle=black, sharp corners=southwest, sharp corners=northeast]
\textit{And at the same time, you should keep the newly generated sentences: 1. PPL similar to the original text. 2. Semantics similar to the original text. 3. The FRE (complexity of the sentence) is similar to the original text.
Hint: Try to find keywords to replace. Avoid complex words and sentences that are not commonly used by humans.}
\end{tcolorbox}

For the RAFT attack, we impose constraints on text quality during word selection. 
These constraints are as follows: while reducing the detection success rate, we strive to maintain the original PPL and FRE as much as possible while also ensuring high semantic similarity. 
Specifically, in our experiment, the constraints are as follows: the change in PPL and FRE of a single sentence before and after modification should be less than 5\%, and the two semantic similarity metrics should be greater than 0.95.
For the TOBLEND attack, we also impose text quality constraints when selecting the next token. 
Instead of random selection, we rank the tokens generated by the four LLMs based on text quality and then choose the one with the highest quality.
Specifically, we first calculate four new candidate tokens and form four candidate sentences by adding each token to the existing sentence. 
We then compute the semantic similarity between each of the four candidate sentences and the original (unattacked) sentence, requiring it to be greater than 0.7. 
Afterward, we rank the four candidate sentences by the difference in PPL between each candidate and the original sentence and select the one with the smallest difference.
We show the persuade code of QPA in RAFT and TOBLEND in~\Cref{code:raft,code:toblend}.
We evaluate the impact of integrating QPA into three evading attacks on the MGT-Academic dataset. 
The results of the text quality assessment are presented in~\Cref{tab:table3}. 
We observe that QPA significantly enhances both semantic similarity metrics across all three attack types. 
Additionally, in most cases, QPA also contributes to improvements in PPL and FRE. 
On the other hand, the experimental results(\Cref{fig:QPA}) also indicate that integrating QPA into evading attacks does not compromise their attack effectiveness.
Furthermore, regarding computational cost, for GPU consumption, the three methods show no significant change before and after adding the QPA module. 
In terms of time consumption, we calculated the averages across three datasets. Prompt’s time increased from 22.803 seconds to 23.231 seconds, RAFT’s from 572.767 seconds to 592.681 seconds, and TOBLEND’s from 25.699 seconds to 30.301 seconds. 
These increases are minimal, suggesting that the integration of QPA introduces only a small overhead in computational cost.

QPA’s core insight can be further refined to enhance its effectiveness by improving both the evaluation of text quality and the dynamic adjustment of attack strategies.
Currently, QPA uses general metrics like PPL, FRE, and semantic similarity, but incorporating more advanced linguistic features such as syntactic fluency, coherence, and context-aware semantic analysis could provide finer control over text quality. 
In addition, a more dynamic approach could involve training a classifier to predict which parts of a sentence most influence detection or semantic similarity, enabling the optimization of attack strategies on a more granular level.
This classifier-based framework could also allow for real-time adjustments to the intensity of attacks, ensuring a more adaptive and context-sensitive approach that better improves the trade-off.

\begin{algorithm}
\caption{RAFT Attack with PQA}
\begin{algorithmic}[1]
\label{code:raft}
\STATE \textbf{Input:} $S = \{w_1, w_2, \dots, w_n\}$, $W = \{w_1, w_2, \dots, w_n\}$
\STATE \textbf{Output:} $S'$
\FOR{each $w_i \in W$}
    \STATE $C_i = \text{LLM}(w_i)$ \quad \text{(Generate candidates for $w_i$)}
    \STATE \textbf{\textbf{QPA:}} \quad \text{Rank $C_i$ by:}
    \STATE $\text{PPL}(C_i, S) - \text{PPL}(S) < 5\%$
    \STATE $\text{FRE}(C_i, S) - \text{FRE}(S) < 5\%$
    \STATE $sim(C_i, S) > 0.95$
    \STATE \text{Select optimal candidate} $c_{i*} \in C_i$
    \STATE Replace $w_i$ with $c_{i*}$
\ENDFOR
\STATE \textbf{Return:} $S'$
\end{algorithmic}
\end{algorithm}

\begin{algorithm}
\caption{TOBLEND Attack with PQA}
\begin{algorithmic}[1]
\label{code:toblend}
\STATE \textbf{Input:} $S = \{w_1, w_2, \dots, w_n\}$
\STATE \textbf{Output:} $S'$
\STATE $O_1, O_2, O_3, O_4 = \text{LLM}(S)$ \quad \text{(Generate outputs from 4 LLMs)}
\FOR{each $t_i \in \{O_1, O_2, O_3, O_4\}$}
    \STATE $S_i = S + t_i$ \quad \text{(Generate candidates by appending token $t_i$)}
    \STATE \textbf{\textbf{QPA:}} \quad \text{Rank $S_i$ by:}
    \STATE $\text{sim}(S_i, S) > 0.7$
    \STATE $\text{PPL}(S_i) - \text{PPL}(S) < 5\%$
    \STATE $\text{FRE}(S_i) - \text{FRE}(S) < 5\%$
    \STATE \text{Select optimal candidate} $S_{i*} \in S_i$
    \STATE Replace $S$ with $S_{i*}$
\ENDFOR
\STATE \textbf{Return:} $S'$
\end{algorithmic}
\end{algorithm}

\section{Attack Blending}
\label{sec:ab}

To validate Attack Blending, we conduct experiments on the WP dataset, using a combination of Dipper and RAFT. 
In this experiment, we apply the two methods alternately, one sentence at a time. 
In our experiment setup, we alternated between using Dipper and RAFT attack methods, applying one method to each sentence in the text. 
The results are shown in~\Cref{tab:blending}.
While this alternating approach demonstrates the effectiveness of Attack Blending, it weakens the full potential of this strategy, as it does not optimally match attack methods to the specific needs of each segment. 
This simplified setup is intended to showcase the foundational effectiveness of Attack Blending, but it is far from an optimal implementation. 
There are many avenues for further improvement, such as training a classifier to evaluate each sentence’s impact on detection effectiveness and text quality.
Such an approach would allow for a more targeted selection of attack methods, optimizing both attack performance and text preservation. 
This highlights the broader potential of Attack Blending, which can be refined and expanded in future research to achieve even better results.

\end{document}